\documentclass{aa}
\usepackage[varg]{txfonts}

\usepackage{natbib}
\bibpunct{(}{)}{;}{a}{}{,} 
\usepackage{graphicx}	
\usepackage{amsmath}	

\usepackage[hyperfootnotes=false, linktocpage=true, breaklinks=true, colorlinks=true, linkcolor=blue, citecolor=blue, urlcolor=blue]{hyperref}

\usepackage{longtable}
\usepackage{xcolor}

\usepackage{booktabs}

\newcommand{\cmsq}{~cm$^{-2}$ }

\def\xmm{{\it XMM-Newton}}

\def\chandra{{\it Chandra}}

\def\herschel{{\it Herschel}}

\begin{document}

\title{Disentangling the Galactic centre X-ray reflection signal using XMM-Newton data} 

\author{K.~Anastasopoulou\inst{1,2}\thanks{E-mail: konstantina.anastasopoulou@cfa.harvard.edu}
\and I.~Khabibullin\inst{3,4,5}
\and E.~Churazov\inst{4,5}
\and G.~Ponti\inst{6,7,8}
\and M. C.~Sormani\inst{8}
\and R. A. Sunyaev\inst{4,5}
\and C.~Maitra\inst{7}
\and S.~Piscitelli\inst{6,8}
}

\institute{Center for Astrophysics $|$ Harvard \& Smithsonian, 60 Garden Street, Cambridge, MA 02138, USA 
\and INAF-Osservatorio Astronomico di Palermo, Piazza del Parlamento 1, 90134 Palermo, Italy
\and Universit\"ats-Sternwarte, Fakult\"at f\"ur  Physik, Ludwig-Maximilians Universität, Scheinerstr. 1, 81679 M\"unchen, Germany
\and  Max Planck Institute for Astrophysics, Karl-Schwarzschild-Str. 1, D-85741 Garching, Germany
\and   Space Research Institute (IKI), Profsoyuznaya 84/32, Moscow 117997, Russia
\and INAF-Osservatorio Astronomico di Brera, Via E. Bianchi 46, I-23807 Merate (LC), Italy
\and  Max-Planck-Institut für extraterrestrische Physik, Gie{\ss}enbachstra{\ss}e 1, D-85748, Garching, Germany
\and Como Lake Center for Astrophysics (CLAP), DiSAT, Università degli Studi dell’Insubria, via Valleggio 11, I-22100 Como, Italy}

\date{Received xxxx / Accepted xxxx}

\abstract{}

 \abstract
   {}    
   {We investigate the X-ray emission from the Galactic Centre (GC) region, focusing on the 6.4\,keV fluorescent line of neutral or weakly ionised iron, which is commonly attributed to X-ray reflection from dense molecular clouds. Our goal is to separate the reflection signal from other physical X-ray components. We aim to produce a clean map of the 6.4\,keV emission, thus providing a better understanding of the X-ray reflection processes in the GC.}
   {We utilised a deep mosaic of all available \xmm{} observations, encompassing the central 40 square degrees of the Galaxy. This dataset integrates information from 503 individual observations, resulting in a total clean exposure time of 7.5\,Ms.
The mosaics of two narrow bands centred at 6.7\,keV and 6.4\,keV, and a broader continuum band at lower energies (5-6.1\,keV), provided valuable spatial and spectral information on the X-ray emission. These combined with the stellar mass distribution of our Galaxy enabled us to decompose the observed signal into physically meaningful components.
   }
   { Our analysis shows that the cleaned 6.4\,keV band map, free from the contribution of bright and unresolved point sources, is predominantly shaped by X-ray reflection from dense molecular clouds. The spatial distribution of this emission, which strongly correlates with the molecular gas distribution in the Central Molecular Zone (CMZ), supports the interpretation that this map provides the best estimate of the X-ray reflection signal averaged over the last two decades. 
   The cleaned reflection map produced could serve as a tool for future studies to quantify upper limits on the reflection contribution from low-energy cosmic rays in unilluminated regions.
    Moreover, we estimate that, on average within the CMZ, approximately 65\% of the ridge emission contributes to the observed emission in the 6.4\,keV band, a factor that should be incorporated into upcoming investigations of the GC such as polarisation studies of the reflected X-ray continuum from molecular clouds and statistical assessments of the reflection surface brightness.}
  {}
\keywords{ISM: clouds -- Galaxy: bulge -- Galaxy: centre -- Galaxy: disc -- X-rays: general -- X-rays: ISM}
\authorrunning{K. Anastasopoulou et al.}
\titlerunning{X-ray reflection in the GC} 
\maketitle



\section{Introduction}

The Galactic centre (GC), the central few degrees of our Galaxy, is an environment of very energetic phenomena (i.e. flaring activity of  Sgr\,A$^{*}$, numerous supernova explosions, stellar winds of massive stars, active stars and various types of X-rays binaries) which result in the emission of soft ($<$4\,keV) as well as hard X-rays \citep[$>$4\,keV; for reviews][]{ponti13,koyama18}. This environment makes the Galactic Centre region rich in (apparently) diffuse X-ray emission. In particular, in hard X-rays, two components are commonly invoked to describe spectral and spatial properties of the emission: the Galactic Ridge X-ray emission and the reflected X-ray emission from the massive molecular clouds.

The Galactic ridge X-ray emission has been so far traced by
the $\sim$6.7\,keV line which originates from He-like iron (Fe XXV) and points to the existence of very hot gas \citep[3-10\,keV; e.g.][]{cooke69,worrall82,koyama86, koyama07b,yamauchi93,revnivtsev09}.
If this hot plasma is bound to compact objects, the spatial distribution of the ridge X-ray emission is expected to follow the smooth distribution of the stellar mass along our Galaxy, including separate contributions of the Galactic Disc and bar, the nuclear stellar disc and nuclear star cluster.
Its spectral shape is relatively well established in regions outside the GC and in the hard X-ray band up to $\sim$60\,keV \citep[e.g.][]{krivonos07,yuasa12,krivonos25}. However, in the very central two degrees of our Galaxy the origin of the 6.7\,keV emission has been widely debated as it was found in excess of what was expected by the stellar mass distribution of the Galaxy. This excess emission is attributed to either unresolved point sources (as a result of a new population or larger population of known systems), real hot plasma (due to past activity of Sgr\,A$^{*}$, or past supernova explosions), or a combination of the two \citep[e.g.][]{koyama89,yamauchi93,muno04,park04,revnivtsev07,uchiyama11,nishiyama13,heard13sources}. 
Therefore, in the central few degrees of the Galaxy, the spectrum of the Galactic ridge X-ray emission remains uncertain, due to the challenge of disentangling diffuse emission from unresolved point sources and the fact that it likely arises from a spatially variable mixture of populations of point sources \citep[accreting white dwarfs and stars; e.g.][]{revnivtsev06b,revnivtsev09,xu16}. 
In a recent study, \citet{anastasopoulou23} found that the observed GC X-ray excess might be the result of  higher X-ray emissivity  per unit stellar mass sources in the GC possibly owing to higher metallicity in the GC \citep[e.g.][]{yamauchi16,feldmeierkrause17,zhu18,do18,schulteis21,fritz21}.
\citet{anastasopoulou23} found a remarkable similarity between the X-ray emission and the stellar mass density distribution in the GC when scaling for the higher X-ray emissivity in the nuclear stellar disc and cluster, while a small remaining excess in the  central 0.3 degrees of the GC can be explained by the contribution of supernova remnants based on the measured supernova rate.

The spectrum of the reflected X-ray emission can be readily predicted, dominated by a $\sim$6.4\,keV line originating from neutral or weakly ionised iron (Fe K$\alpha$), and points to the existence of cold gas. However, its spatial distribution is rather complicated.
In the central degrees of the GC, the region known as the central molecular zone (CMZ), the emission from  the 6.4\,keV line has been found to be asymmetric (in contrast with the more or less uniform distribution of the Galactic ridge X-ray  emission), and be correlated spatially with GC molecular clouds which are also highly asymmetrically distributed, with approximately three-quarters found at positive longitudes \citep{bally88}. The main mechanism of its production is considered to be X-ray reflection off dense molecular clouds which act as mirrors to the past flaring  activity of the central supermassive black hole Sgr\,A$^{*}$ on timescales of hundreds of years \citep[e.g.][]{sunyaev93,koyama96,ponti13, churazov17,khabibullin22}.  
Studying in detail the X-ray reflection from molecular clouds has significantly enhanced our understanding of the type/position of the illuminating source and its variability \citep[e.g.][]{ponti10,clavel13,terrier10,chuard18,kuznetsova22,stel25}.
In addition to reflection, other mechanisms could account for the observed continuum and the 6.4\,keV emission line in the CMZ. One such mechanism is collisional ionisation from accelerated particles \citep[cosmic ray electrons, protons/ions; e.g.][]{valinia00, yused-zadeh02, yusef-zadeh07, bykov02,dogiel09}, which has initially been proposed to be the case of emission in molecular gas around Arches cluster as no variability has been observed over an eight year period \citep[e.g.][]{wang06,capelli11}. However, recent studies have shown significant variability around the Arches cluster over a 20-year span, with some residual emission still possibly attributed to cosmic rays \citep[e.g.][]{clavel14,krivonos17,chernyshov18,kuznetsova19,stel25}.
Observations of fast flux variations in the 6.4\,keV emitting clouds
\citep[e.g.][]{muno07,ponti10,capelli12,clavel13,churazov17a,inui09,terrier10,ryu13,chuard18}, and polarisation \citep[][]{churazov17,marin23} argue against a cosmic ray-induced Fe K$\alpha$ origin, at least for the variable component of the emission \citep[i.e. review by][]{ponti13}.

Several studies have concluded that point sources contribute minimally to the 6.4 keV emission in the Galactic Center. 
For instance, \citet{murakami01}, using a 100\,ks \chandra{} observation of the central 3$\times$3.5 arcmin region of the Sgr B2 molecular cloud, resolved approximately 18 sources which account for only 3\% of the luminosity produced by reflection. Moreover, \citet{wang02} using \chandra{} data to map the central degrees of the Galactic Center  extracted composite spectra of the resolved point source population, and confirmed that these sources contribute minimally to the 6.4\,keV emission. Instead, they showed that these sources account for most of the observed 6.7\,keV emission. 
However, very deep 
$\sim$1\,Ms \chandra{} observations of the Galactic ridge and the GC \citep{revnivtsev07, revnivtsev09}, have shown that a substantial portion of the 4–8 keV energy range is contributed by faint sources. Astrophysical objects like low-mass X-ray binaries and various types of cataclysmic variables are known to produce strong 6.4 keV emission lines, primarily from the inner regions of their accretion discs and reflection from the white dwarf surface \citep[e.g.][]{kallman89,barret2000}.

In this work, we aim at combining spatial and spectral information on the X-ray emission provided by the deep \xmm{} mosaic of the Galactic centre and inner Galactic Disc region in order to decompose the signal into physically-motivated components. Namely, using the spatial correlations of the continuum, 6.7\,keV and 6.4\,keV bands, together with the models of the stellar distribution in this region, we derive a "residual" 6.4\,keV emission map, which we argue that offers the best estimate for the X-ray reflection signal in the Galactic centre.
This paper is organised as follows: in Section \ref{data} we describe the \xmm{} observations comprising the big mosaic and the stellar mass distribution models, in Section \ref{method} we present the method we used in order to produce the clean reflection emission maps. In Section \ref{resultsdiscussion} we present and discuss our results and compare with known maps of molecular clouds in the Galactic Centre. Finally, we provide our conclusions in Section \ref{conclusions}.
\section{Data}\label{data}

\subsection{X-ray data}

For this project, we have used all  \xmm{} observations available until 25th of February 2024 (503; $\sim$7.5\,Ms clean -- background flare- and vignetting-corrected -- exposure time EPIC-pn equivalent), with exposure time greater than 5\,ks, within the central 40 square degrees of our Galaxy. The list of observations is reported in \citet{anastasopoulou23}, \citet{ponti19}, and \citet{ponti15}. 
The information of the most recent 133 observations are reported in Table\,\ref{tab.xmm} in the Appendix.

\begin{figure}[!htbp]
\centering
\includegraphics[angle=0,trim=1cm 5.5cm 1cm 2.5cm,width=0.99\columnwidth]{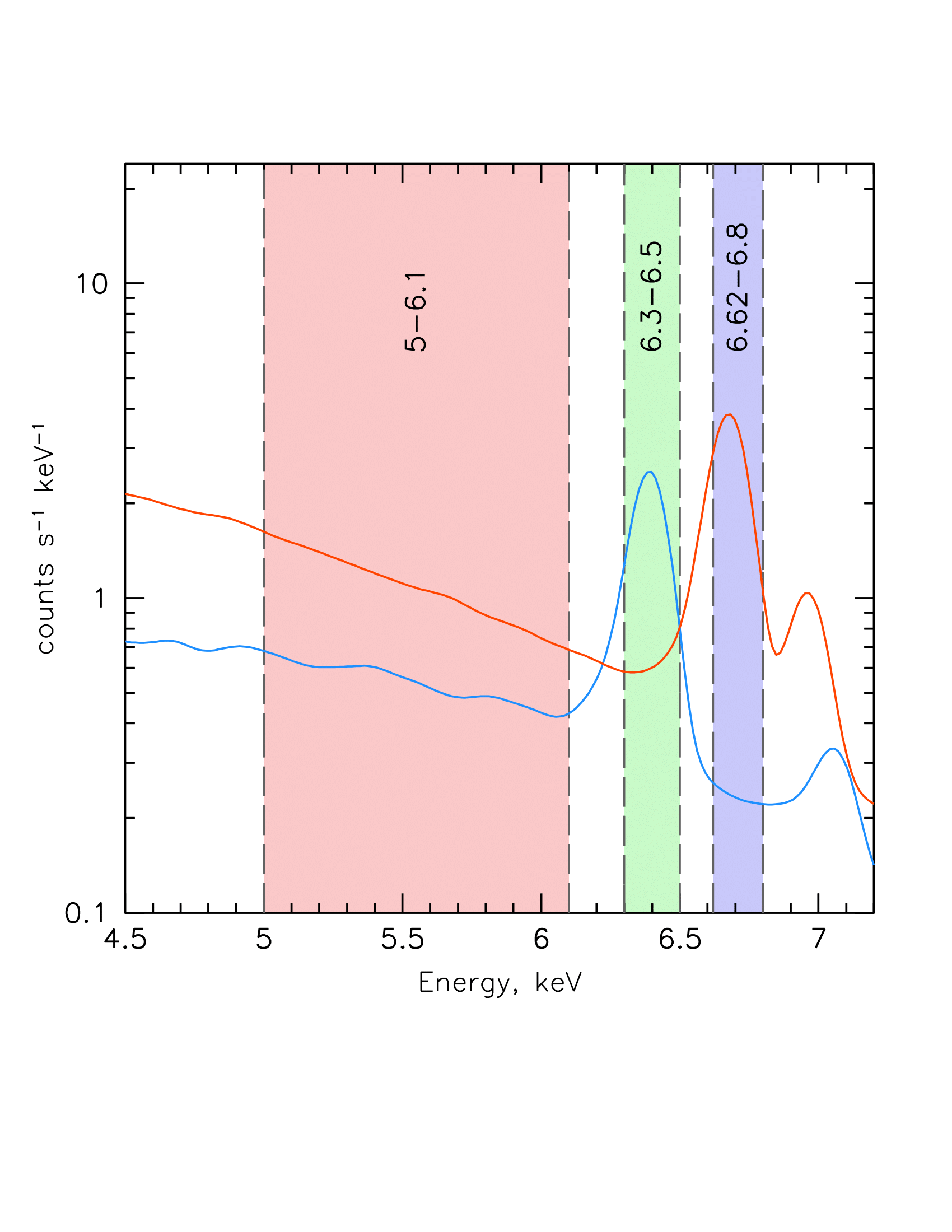}
\caption{Spectral representation of the three energy bands used through this project. The bands are highlighted by the shaded regions. 
The representative model spectra (convolved with the \xmm{} spectra response function) of hot plasma (\texttt{apec} model in XSPEC) and reflection (\texttt{crefl16} model; \citealt{churazov17a}) are shown with the orange and blue lines respectively.} 
\label{fig:bands}
\end{figure}

\begin{figure*}[!htbp]
\centering
\includegraphics[angle=0,trim=0cm 0cm 0cm 0cm,width=1.99\columnwidth]{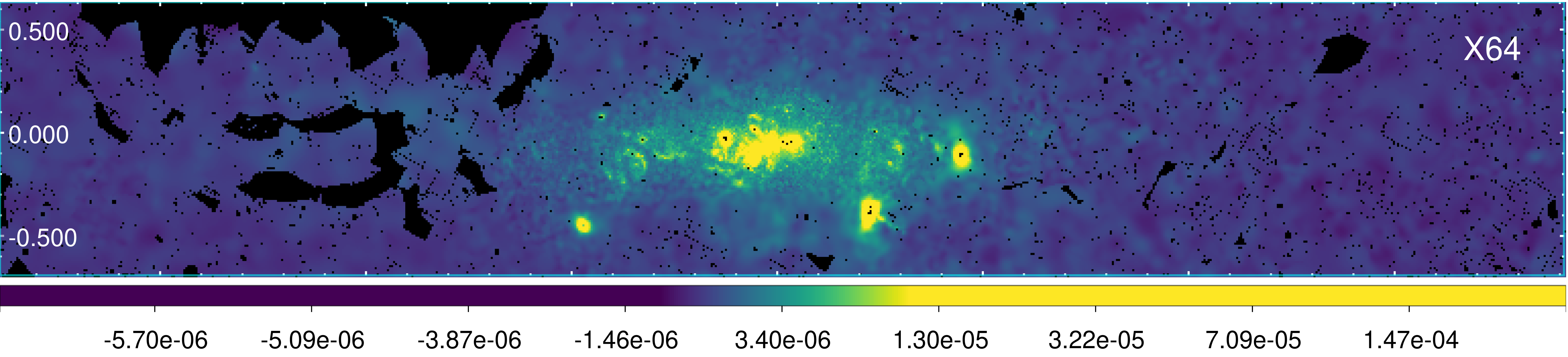}
\includegraphics[angle=0,trim=0cm 0cm 0cm 0cm,width=1.99\columnwidth]{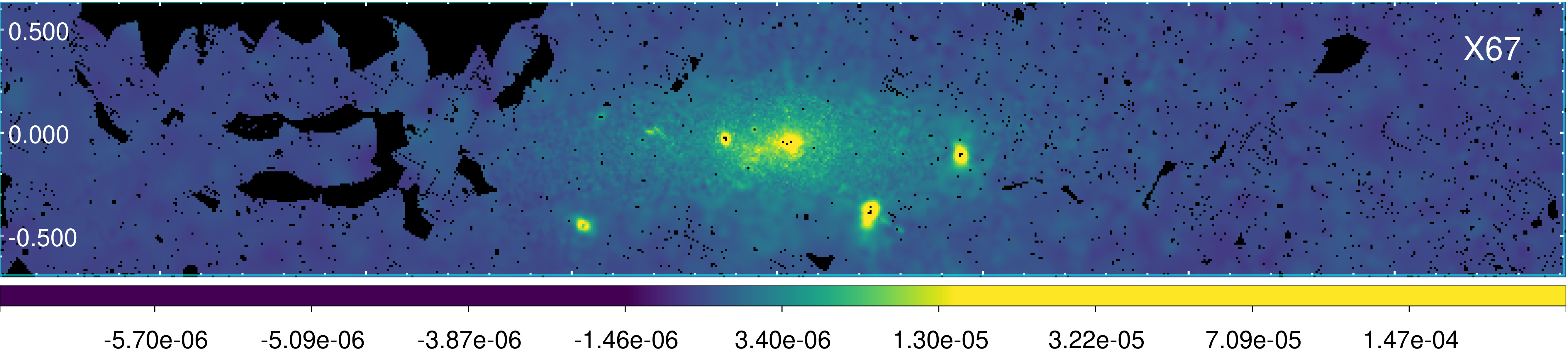}
\includegraphics[angle=0,trim=0cm 0cm 0cm 0cm,width=1.99\columnwidth]{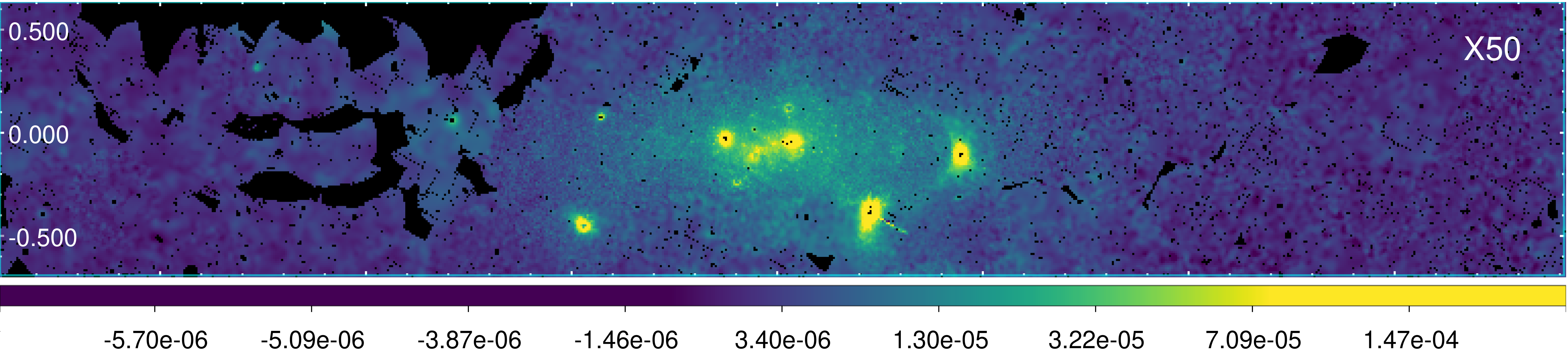}
\includegraphics[angle=0,trim=0cm 0cm 0cm 0cm,width=1.99\columnwidth]{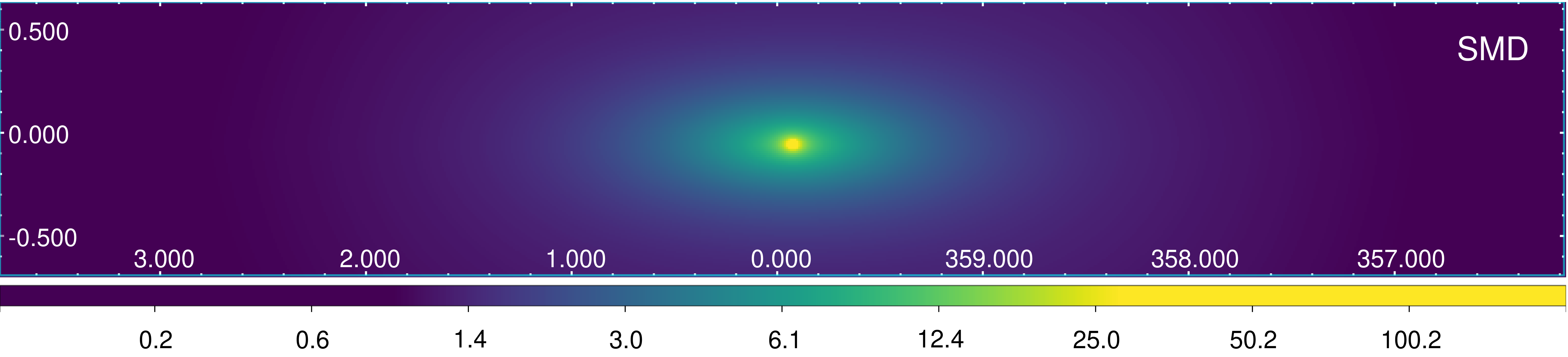}
\caption{Raw maps used throughout this study. From top to bottom: the \xmm{} 6.4\,keV (X64),  6.7\,keV (X67), the continuum band (X50), and the stellar mass distribution (SMD). The \xmm{} maps are measured in count per second per pixel (cr/pix), whereas the SMD map represents stellar density in $10^3$M$_{\odot}$/pix. The colour bar ranges are selected arbitrarily to emphasize and display morphological differences. The x-axis and y-axis represent the Galactic longitude and latitude, respectively, in degrees.} 
\label{fig:maps}
\end{figure*}

The calibration and analysis of the observations was performed using the standard tools of the \xmm{} Science analysis system (SAS) v21.0.0, and is reported in detail in \citet{anastasopoulou23}. The final product of the analysis is a stray-light free, background subtracted, source excised, and adaptively smoothed mosaic for each \xmm{} detector, and each energy band of interest. The MOS and pn mosaics were combined accounting for the effective area differences between the detectors at the corresponding energy band \citep[details available in the appendix of][]{anastasopoulou23}. To isolate the fluorescent Fe K emission from other contaminating processes in the Galactic Center, we produced mosaics for three \xmm{} energy bands, $6.3-6.5$\,keV (X64), $6.62-6.8$\,keV (X67), and $5.0-6.1$\,keV (X50). The mosaics in this work are always presented in units of counts per second per pixel (cr/pix), with each pixel measuring 8$\times$8 arcseconds. These bands represent the Fe~XXV from very hot plasma, the neutral or weakly ionised Fe K$\alpha$ fluorescent emission, and the continuum at lower energies respectively. 
The spectral shape of each component is shown in Fig\,\ref{fig:bands} where the X67 band is intentionally slightly offset from the 6.7\,keV line peak to minimize contamination from the nearby 6.4\,keV fluorescent line. The central regions ($\sim$8$\times$1 degrees) of the raw mosaics at each band are presented in the first three panels of Fig.\,\ref{fig:maps}.

\subsection{Stellar mass distribution model}\label{smd}

The total stellar density of the Milky Way can be represented as a sum of distinct components:

\begin{equation}
\rho_{\rm TOT}(x,y,z) = \rho_{\rm NSC} + \rho_{\rm NSD} + \rho_{\rm BAR} + \rho_{\rm DISC} \,.
\end{equation}

These components, arranged by increasing Galactocentric radius $R$, include: 
\begin{itemize}
    \item  the nuclear star cluster (NSC), a dense and slightly flattened structure of stars, surrounding Sgr\,A$^{*}$  with a mass  $\approx 2.5 \times 10^7 \, M_{\odot}$, dominating within $R \lesssim 10 \, \text{pc}$ \citep{schoedel14, neumayer20}. 
    \item  the nuclear stellar disc (NSD), a flattened distribution with a mass of approximately $1.05 \times 10^9 \, M_{\odot}$, prevalent at $10 \lesssim R \lesssim 200 \, \text{pc}$ \citep{launhardt02, sormani22nsd}.
    \item the Galactic bar, an elongated structure in the Galactic plane with a mass near $1.9 \times 10^{10} \, M_{\odot}$, which dominates within $0.2 \lesssim R \lesssim 3 \, \text{kpc}$ \citep{Bland-Hawthorn2016}.
    \item the Galactic Disc, which is the main contributor to the stellar mass density beyond $R \gtrsim 3 \, \text{kpc}$.
\end{itemize}

The stellar mass distribution model of our Galaxy is created by combining recent models of each individual component. For this project we use the components described as Model 2 in table 2 and section 2.2 of \citet{anastasopoulou23}. Namely, we used, for the NSC the best-fitting model from \citet{chatzopoulos15}\footnote{The model by \citet{feldmeierkrause17}  estimates a lower total mass for the NSC.}, for the NSD the fiducial model of \citet{sormani22nsd}, for the Galactic bar the bar + long bar from \citet{sormani22bar} which is based on the numerical models by \citet{portail17}, and for the Galactic Disc the disc from \citet{sormani22bar}. The stellar mass distribution model, calculated by integrating the volume density ($\rho(x,y,z)$) along the line of sight to obtain the surface density on the plane of the sky ($\Sigma (l,b)$ where $l$ and $b$ are the Galactic coordinates), is shown in the bottom panel of Fig.\,\ref{fig:maps}.

\section{Method}\label{method}
The Galactic Plane (GP) is densely populated with many different objects (and classes of objects). It is therefore not surprising that from any location in the GP, we almost always see a composite X-ray spectrum that includes contributions of many unrelated objects along the same line of sight. There are many ways of emphasizing one particular component. These range from selecting an energy band, where this component is expected to dominate, to fitting a multi-component spectral model for each sky position. The former is very simple but usually results in a map that can be severely contaminated by other components, while the latter approach might be too complicated and lead to noisy results if the spectral model is non-linear. One can partially mitigate the noise problem by resorting to linear spectral models \citep[see example in][for a two-components case]{2017MNRAS.465...45C}, which works well if the spectral shapes of different components are known. In this case,  their normalizations can be independently determined at every position via a linear combination of data. A similar approach can be applied to a collection of maps as is often done in the analysis of Cosmic Microwave Background (CMB) data \citep[e.g.][]{2016A&A...594A..22P}. For a known spectral shape, one can predict the contribution of the component $j$ with amplitude $A_j(x,y)$ to any map $i$, i.e. $m_i(x,y)=\Sigma_j A_j(x,y) s_{j,i}$. For instance, if $A_j(x,y)$ is a surface density of sources of type $j$, e.g., stars, their contribution to the X-ray map $m_i(x,y)$ in the band $i$ is set by the flux $s_{j,i}$ such sources produce in this band. Therefore, knowing $s_{j,i}$ one can select the coefficients $\alpha_i$ so that a linear combination of maps $m(x,y)=\Sigma_i \alpha_i m_i(x,y)$  will eliminate some of the spectral components exactly and preserve the normalization of the component of interest (at least for the noise-free data).  A more practical version of the same procedure is to keep only the requirement of preserving the normalization of one spectral component and then choose the coefficients so that the resulting map has the smallest L2 norm $\int m^2(x,y)dxdy$.  If the component of interest is confined to one particular band/map, i.e. $s_{0,i}=0$ for $i\neq 0$, the above procedure reduces to finding the coefficient $\alpha_i$ that minimize the norm:
\begin{eqnarray}
\min_{\alpha} \Sigma_{x,y}\left [ \frac{ m_0(x,y)-\Sigma_i\alpha_im_i(x,y)}{\Sigma_0(x,y)}\right ]^2, \nonumber
\end{eqnarray}

where $m_0(x,y)$ is the reference map containing the component of interest (and the contributions of other components too), $\Sigma^2_0(x,y)$ is the estimated variance at a given position of the reference map (in case if it is known and other maps are free from the noise), and $m_i(x,y)\,\ {\rm for}\,i=1,N$ are the all other available maps. In this case, the maps need not be X-ray images but could be any images from other bands or models.

The shortcomings and limitations of the above approach are obvious. However, if strong spatial variations of different components are present, this procedure could reveal the level of correlations between the maps and, potentially, clean the maps from some of the contaminating signals. 
The method we use, compared to equivalent width (EW) approaches, is not biased by regions with low stellar mass distribution, which can artificially increase the EW due to reduced continuum flux. Our method is less sensitive to variations in the SMD and provides a more accurate representation of the clean 6.4 keV emission, enabling direct comparison with the dense gas distribution in the GC.
In what follows we make another simplifying assumption by setting $\Sigma_0(x,y)=1$. Effectively, this means that the algorithm minimises the variance of the residual image after removing the contributions of all components. This is a reasonable choice since we are dealing with smoothed images and mask regions heavily affected by systematics. 

Our set of maps, presented in the previous sections, is summarized in Table~\ref{tab:maps}. This study is focused on the X-ray map X64. This map contains the fluorescent line of neutral (or weakly ionized) iron. Other maps serve as possible proxies for various components, which may spatially correlate with our primary maps potentially revealing the origin of the emission \citep{anastasopoulou23}.
We utilize two distinct maps to represent the stellar mass distribution across our Galaxy. The first, labelled N, includes contributions from the NSC and the NSD. The second map, labelled B, accounts for contributions from the Galactic bar and the Galactic Disc. These components (N and B) are analysed separately to account for differences in X-ray emissivity between them, as highlighted in \citet{anastasopoulou23}.

\begin{table}[!htbp]
\centering
\caption{Set of maps used in the analysis.}
\begin{tabular}{lrr}
\toprule
Map & Type & Band    \\ 
\midrule
X64\tablefootmark{a} & X-rays & 6.30--6.50 keV  \\
X67\tablefootmark{b} & X-rays & 6.62--6.80 keV  \\
X50\tablefootmark{c} & X-rays & 5.0--6.1 keV   \\
N\tablefootmark{d} & Model & based on kinematic data  \\
B\tablefootmark{e} & Model & based on kinematic data  \\
Const\tablefootmark{f} & Model & \\
\bottomrule
\end{tabular}
\tablefoot{\tablefoottext{a}{Dominated by 6.4 keV line (proxy to reflected emission)}
\tablefoottext{b}{Dominated by 6.7 keV line  (proxy to hot plasma)}
\tablefoottext{c}{ Continuum below the fluorescent line}
\tablefoottext{d}{Projected distribution of stars in the NSC and NSD}
\tablefoottext{e}{Projected distribution of stars in the Galactic bar and Disc}
\tablefoottext{f}{Spatially flat component}
}
\label{tab:maps}
\end{table}

\subsection*{Inclusion and exclusion regions}

\begin{figure*}[!htbp]
\centering
\includegraphics[angle=0,trim=0cm 0cm 0cm 0cm,width=1.99\columnwidth]{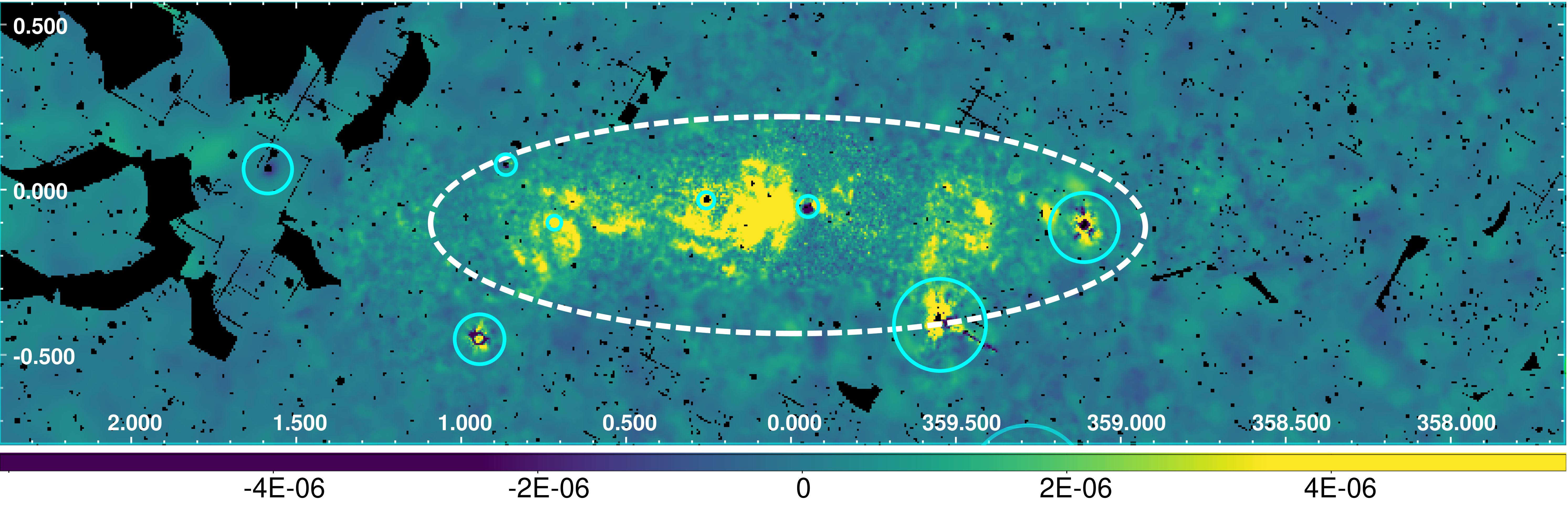}
\caption{Example of the cleaned X64 band map (X64 minus C-X67-X50), with the mask shown as a dashed ellipse and solid circles corresponding to bright sources. The map is  provided in units of counts per second per pixel while black regions indicate areas with no exposure, which may result from the removal of point sources, stray-light artifacts, or chip gaps. } 
\label{fig.examples}
\end{figure*}

Our primary goal is to study the origin of the (apparently) diffuse emission by studying its spatial correlation with other components. From this point of view, the emission from bright X-ray sources is the source of contamination. While the bright point sources have already been excluded during the map-making process, some residual traces are present most probably due to dust scattering halos around these sources \citep[e.g.][]{jin2017,jin18}. For that reason, we applied an additional mask containing a list of regions (Table\,\ref{tab.extrasources}) including bright sources located within these areas. Part of the bright-source mask is shown in Fig.\,\ref{fig.examples} with cyan circles. 

\begin{table}[!htbp]
	\centering
	\caption{Masked regions corresponding to X-ray bright sources}
		\begin{tabular}{@{}lrrr@{}}
			\toprule
			Included bright source  & $\ell$  & $b$  & radius \\
			 & deg & deg  & arcmin \\
			 \midrule
                IGR J17252--3616 & 351.49 & -0.37 & 5.91 \\
              multiple  & 353.56 & -0.72 & 16.69 \\
             multiple& 354.27 & -0.15 & 15.40 \\   
            multiple & 358.56 & -2.21 & 18.90 \\
             1E 1740.7--2942 & 359.11 & -0.11 & 6.20 \\
			SLX 1744--299/300& 359.28& -0.89 & 10.83\\
			2E 1742.9--2929 & 359.54 & -0.41 & 8.40 \\
		    Sgr A East & 359.94 & -0.05 & 1.92 
		    \\    
            1E 1743.1--2843 & 0.26 & -0.03 &1.49 \\
            V* V2606 Oph & 0.67 & 1.18 & 5.76 \\
           CXOGC J174742.4-282227 & 0.72 & -0.10 & 1.28 \\
           multiple & 0.81 & -2.60 & 15.63 \\
		    G0.9+0.1 & 0.87 & 0.08 & 1.88 \\    
		    IGR J17497--2821&0.94 & -0.45 & 4.50 \\	
          HD 161103  & 1.39 & 1.07 & 14.78 \\
             AX J1749.1--2733   & 1.59 & 0.06 & 4.42 \\
              multiple & 4.78 & -0.73 & 18.81 \\
       multiple  & 5.12 & 0.65 & 15.16 \\
         multiple & 5.25 & -0.78 & 24.78 \\
       multiple  & 5.52 & 0.64 & 16.04 \\
			\bottomrule		
			\end{tabular} 	
		\label{tab.extrasources}
          \end{table}

Also, as explained in the previous section, the basic assumption is that physically unrelated signals are expected to be spatially uncorrelated. This is likely not true for the Galactic Center region, as the density of all classes of sources increases towards the centre of the Milky Way. To partially mitigate this problem, we additionally calculate all correlation coefficients for the entire extent of the map after masking the central region (ellipse  shown in Fig.~\ref{fig.examples}) that contains much of the gradient in the surface density of stars or other objects. Masking the central regions is particularly important because most of the molecular clouds that can be sources of "reflected" emission are located there (see Section \ref{contaminatigcomponents}). 
In the next sections, we apply this technique to the X64 map obtained by XMM-Newton.

\section{Results and Discussion}\label{resultsdiscussion}

\subsection{Accounting for different contaminating components}\label{contaminatigcomponents}

Using the method outlined in Section\,\ref{method}, we calculated the contributions of all potential components (listed in Table\,\ref{tab:maps}) to the X64 map. This analysis is presented  for the combined masking of the elliptical central region and bright sources, as it yielded significantly better results compared to using the bright source mask alone, which led to prominent negative residuals (briefly discussed in Section\,\ref{brightmask} of the appendix).
In Table\,\ref{tab.coef}, we present the best-fit coefficients for all possible combinations of the model X64$-($C$ +c_1$N$+c_2$B$+c_3$X67$+c_4$X50) and in 
Table\,\ref{tab.rms} we summarise the different component combinations and the corresponding ratio of the final map variance to the initial map variance for the pixel values across the image.  
The variance provides a measure of how much noise is reduced after processing the image, such as masking bright sources or regions. However, it is important to note that real astrophysical features, can also contribute to an increase in the variance. Therefore, both noise reduction and the potential contribution of physical structures must be carefully considered when interpreting variance changes.

\begin{table}[!htbp]
\caption{Best-fit coefficients}
\setlength{\tabcolsep}{3pt} 
\begin{tabular}{lccccc}
\toprule
Set&C& $c_1$ &$c_2$ &$c_3$& $c_4$  \\ 
&$\times10^{-7}$& $\times10^{-7}$ &$\times10^{-7}$ &$\times10^{-1}$& $\times10^{-1}$  \\ 
&cr/pix& $\frac{cr/pix}{10^3M_{\odot}/pix}$    &$\frac{cr/pix}{10^3M_{\odot}/pix}$ & &  \\ 
\midrule
C & 3.78 & 0 & 0 & 0 &0 \\
C-N &3.21&8.78& 0 & 0 & 0\\
C-N-B &1.13&7.15&2.79 &0  & 0\\
C-N-B-X67-X50 & 0.53 & 3.29 & 0.71 & 2.08 & 0.66\\
*C-X67-X50&0.52  & 0 & 0 &3.26   &  0.77 \\
\bottomrule
\end{tabular}
\tablefoot{General model: X64$-($C$ +c_1$N$+c_2$B$+c_3$X67$+c_4$X50). With asterisk we mark the fiducial model. The pixel size is 8$\times8$ arcseconds.}
\label{tab.coef}
\end{table}

\begin{table}[!htbp]
\caption{X64 band modelling}
\begin{center}
\begin{tabular}{clc}
\toprule
Reference Map ($m_0$) & Set  & $\langle m^2 \rangle / \langle m_0^2 \rangle$\\ 
&&$\times10^{-1}$\\
\midrule
X64 & C &  3.86\\
X64 & C-N   &2.67\\
X64 & C-N-B & 2.58\\
X64 & C-N-B-X67-X50   &2.02\\
X64 &  C-X67-X50  & 2.13\\
\bottomrule
\end{tabular}
\end{center}
\label{tab.rms}
\end{table}

Overall, we find that the residuals for the C-N-B model are slightly lower than those for the C-N model. The lowest variance results are obtained by removing either the C-X67-X50 or C-N-B-X67-X50 components (Table\,\ref{tab.rms}). When we apply the best-fit coefficients to the entire map, we obtain no negative residuals, only slight residuals near very bright sources (Fig.\,\ref{fig.examples}).

\begin{figure*}[!htbp]
    \centering
    \includegraphics[width=1.7\columnwidth]{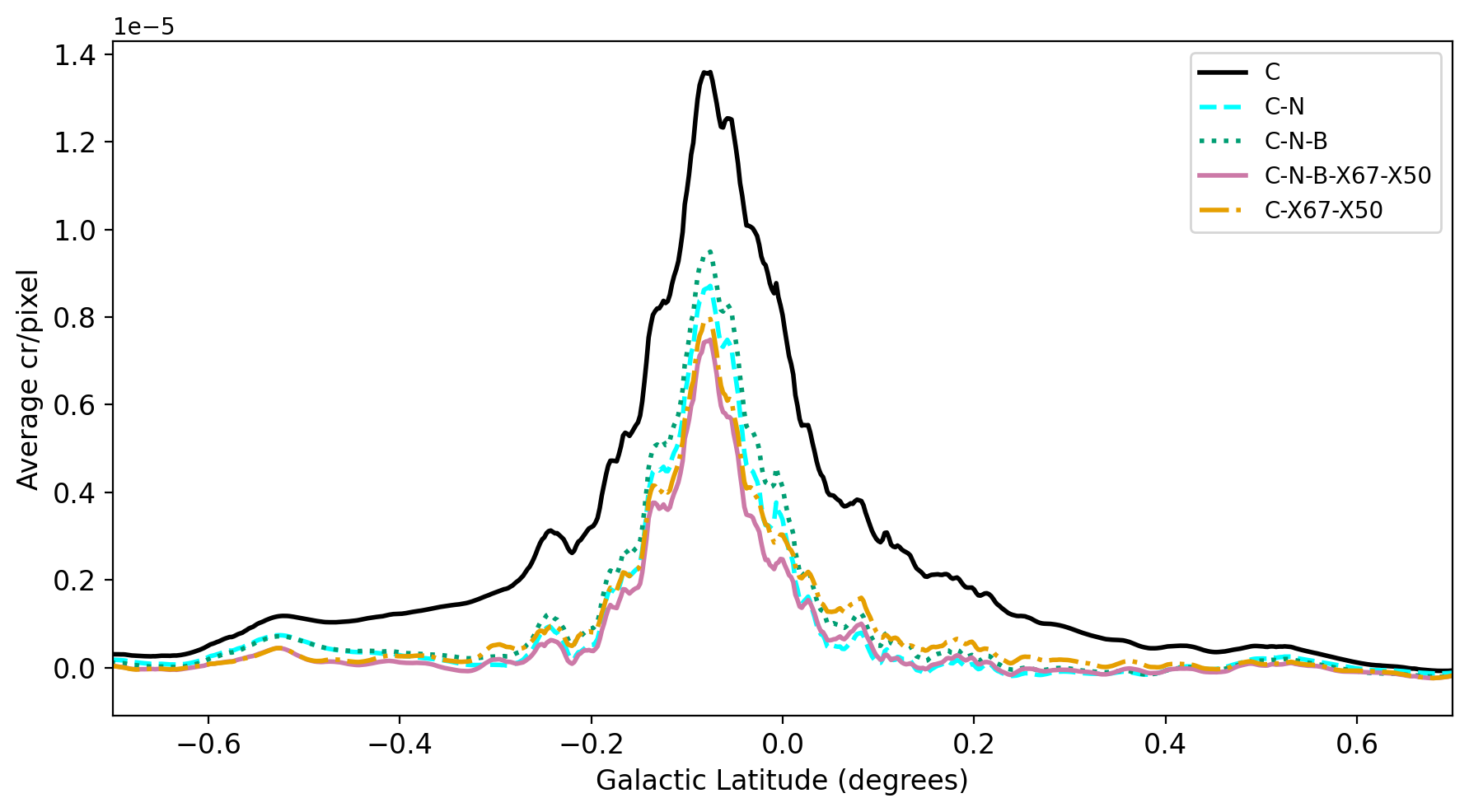}
   \includegraphics[width=1.7\columnwidth]{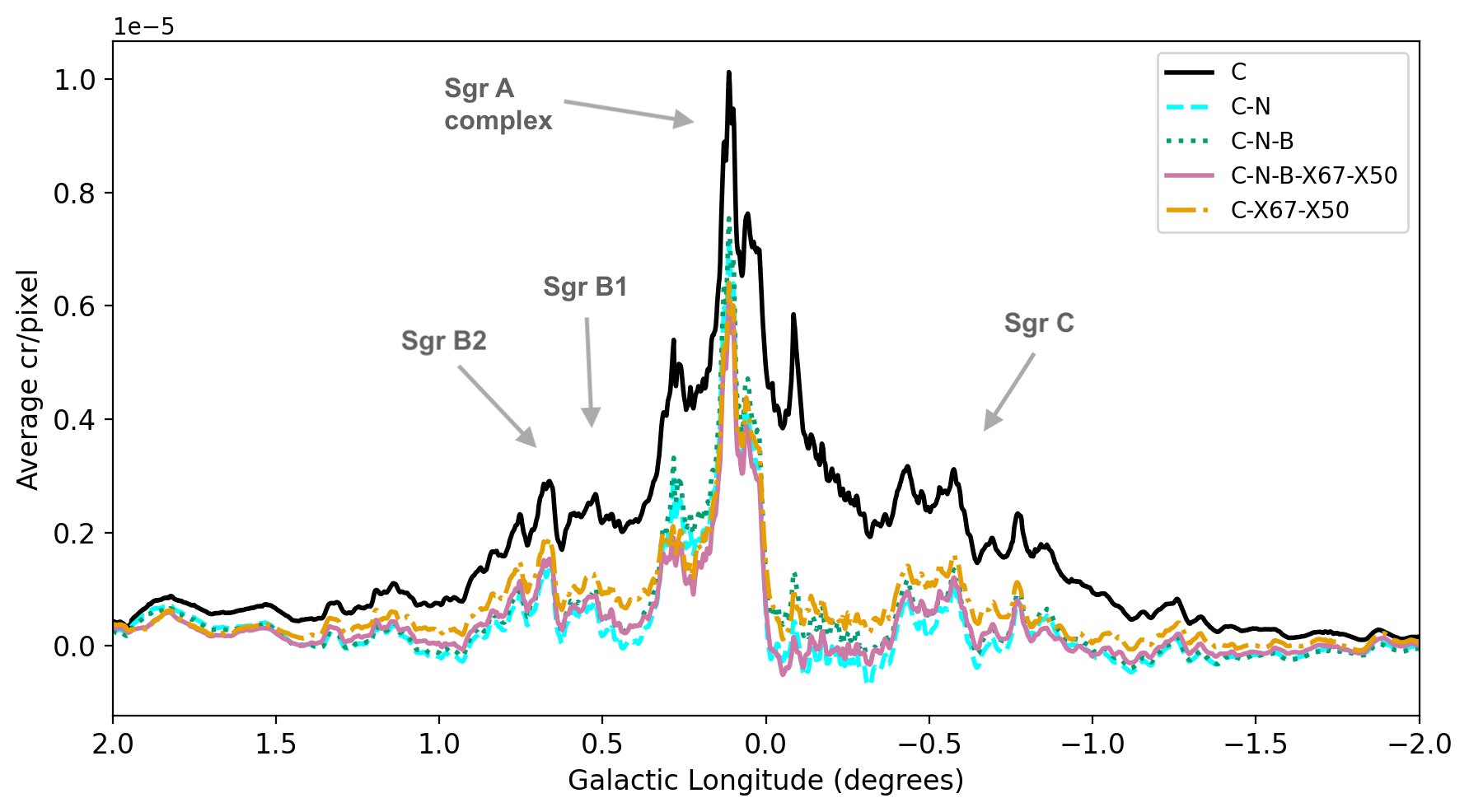}
    \caption{
        Latitudinal (top) and longitudinal (bottom) profiles of the X64 cleaned map. 
        The profiles have a width of 0.5 degrees and the different lines correspond to the different components subtracted from the raw 6.4\,keV \xmm{} map. 
    }
    \label{fig.profiles}
\end{figure*}

To quantify further the differences between the various components, we extracted latitudinal and longitudinal profiles from the corresponding maps (Fig.\,\ref{fig.profiles}).
The profiles represent average values within a 0.5-degree width centred on Sgr\,A$^{*}$. We present the latitudinal profiles along the central 1.2 degrees and longitudinal profiles along  the central 4 degrees. We do not show regions farther from the centre, since there the profiles converge showing emission values close to zero.  
Regions containing bright sources (listed in Table\,\ref{tab.extrasources}) were masked prior to extracting the profiles. Including these regions resulted in numerous bright peaks, complicating the interpretation of the results.
We observe that all models differ slightly, with models C-N (cyan-dashed line)  and and C-N-B-X67-X50 (solid pink line) showing slight negative values particularly along the longitudinal profile (bottom panel 
 of Fig.\,\ref{fig.profiles}). The models with the smaller residual noise are  C-N-B-X67-X50  and C-X67-X50 shown by the pink solid and the yellow dash-dotted lines respectively. Among the two models, we choose model C-X67-X50  as the fiducial one, since it exhibits no negative residuals.

Model C-N-B, represented by the dotted green line in  Fig.\,\ref{fig.profiles}, has an important application, as the cleaned X64 map is obtained solely by subtracting the SMD model contribution. This allows one to use the best-fit coefficients from Table\,\ref{tab.coef} along with the analytical model in Section\,\ref{smd} to estimate the "non-reflection" component contributing to the 6.4\,keV band. 
Following Table\,\ref{tab.coef}, this component is given by (1.13 + 7.15N + 2.79B)$\times$10$^{-7}$ cr/pix,
where the N component, corresponding to the GC region, yields a 6.3--6.5~keV luminosity\footnote{We utilised a distance to the GC of 8.2\,kpc \citep{gravity19}, an EPIC pn effective area of 460\cmsq{}, and a mass surface density of 10$^3M_{\odot}$\,pix$^{-1}$.} per unit stellar mass of 
$L_X[6.3\text{--}6.5, \mathrm{stellar}]\sim1.3 \times 10^{26} \, \mathrm{erg\,s^{-1}\,M_\odot^{-1}}$. 
For comparison, \citet{revnivtsev06a} report a 3--20~keV luminosity per stellar mass of 
$(3.5 \pm 0.5) \times 10^{27} \, \mathrm{erg\,s^{-1}\,}M\mathrm{_\odot^{-1}}$, 
which corresponds to 
$L_X[6.3\text{--}6.5, \mathrm{apec}]\sim(6 \pm 1) \times 10^{25} \, \mathrm{erg\,s^{-1}\,}M\mathrm{_\odot^{-1}}$ 
for a thermal plasma with $kT\sim10$~keV. 
Subtracting this contribution, we estimate the 6.4~keV line luminosity as 
$L_X[6.3\text{--}6.5, \mathrm{line}]\sim7 \times 10^{25} \, \mathrm{erg\,s^{-1}\,}M\mathrm{_\odot^{-1}}$, 
corresponding to an equivalent width of 
$\mathrm{EW}\sim200$~eV, consistent with the spectra of cataclysmic variables from which this emission is thought to originate.
This could provide valuable insights for future studies such as cosmic ray investigations in the GC.
For the C-N-B model, we also observe that the NSC and NSD components (denoted as N) are scaled approximately 2.5 times higher than the bar and Disc components (denoted as B). This behaviour aligns with the finding in \citet{anastasopoulou23}, that these components require different scaling to account for the Galactic Center X-ray excess (estimated at $\sim$1.9).

\subsection{The clean fluorescent map}

\begin{figure*}[!htbp]
\centering
\includegraphics[width=1.9\columnwidth]{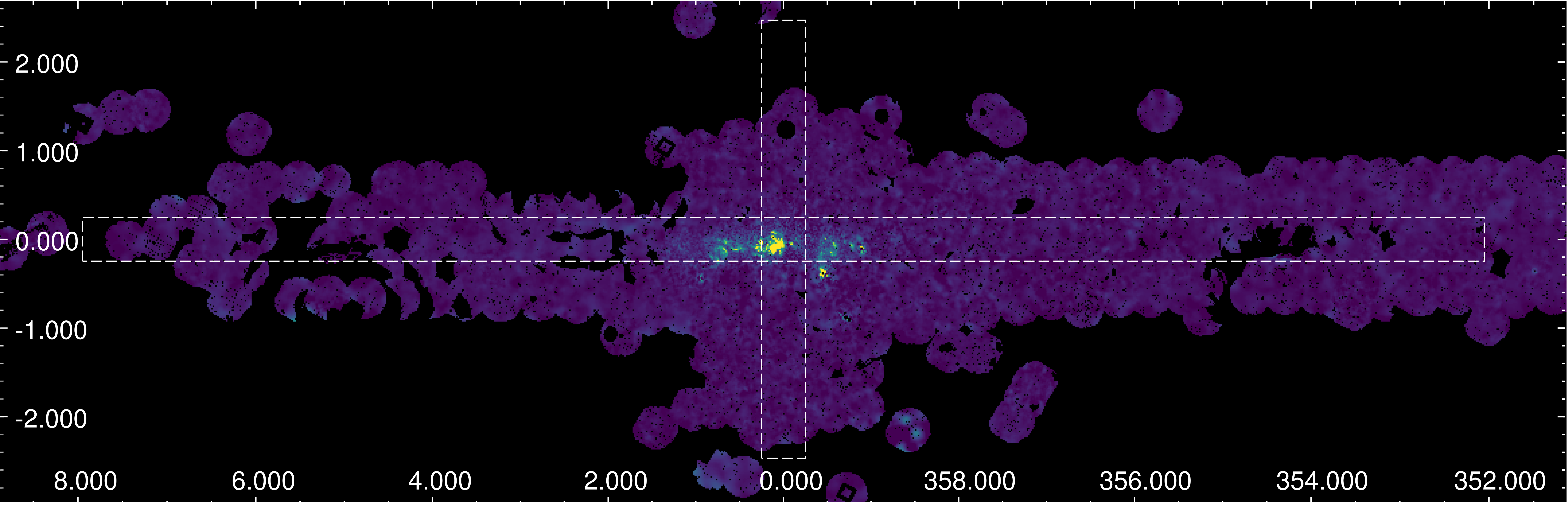}
\includegraphics[width=1.9\columnwidth]{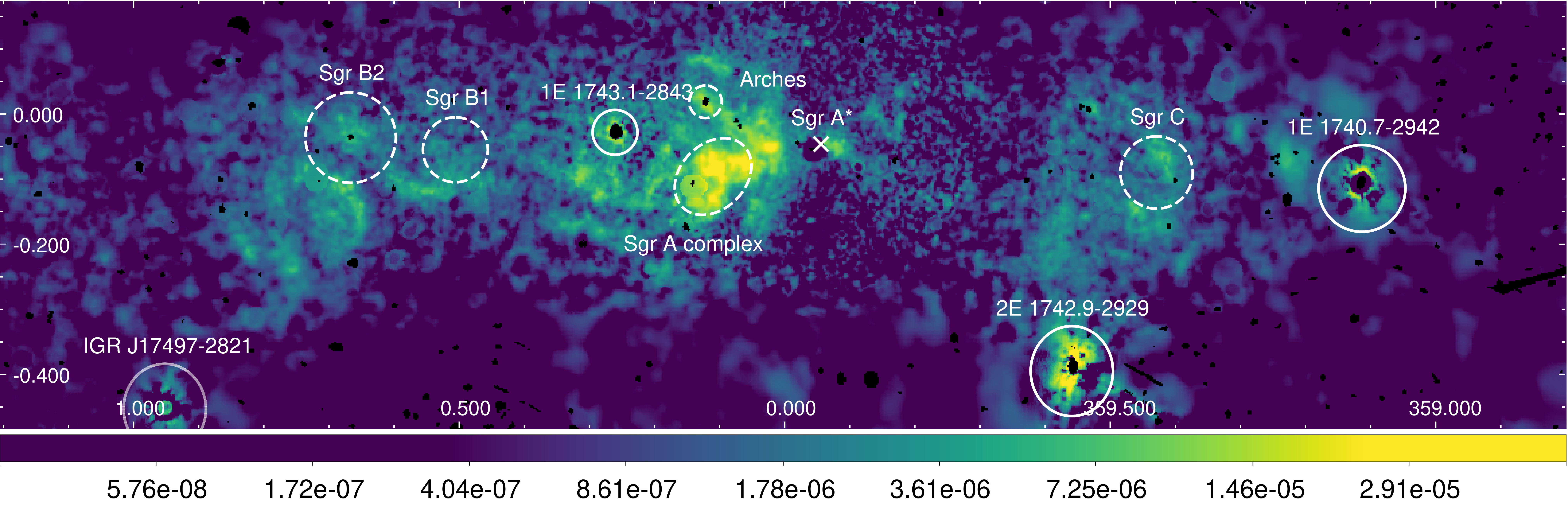}
\caption{X64 cleaned full map and zoomed-in view, averaged over $\sim$20 years of \xmm{} observations. Top: The X-ray emission across the entire X64 cleaned map representing the reflection. The white rectangular regions indicate the areas used for profile extraction. Bottom: A zoomed-in view of the central degrees of the Galaxy, where the reflected emission is concentrated. Bright sources masked out during the analysis are marked with solid white circles, while known molecular clouds are outlined with dashed white regions. The position of Sgr\,A$^{*}$ is denoted by an X.} 
\label{fig.cleanmap}
\end{figure*}

We present the full \xmm{} mosaic of the X64 map cleaned for contributions of other components (model C-X67-X50) in the top panel of Fig.\,\ref{fig.cleanmap} along with the regions used for profile extraction. We assume that this map represents a reflected emission coming from dense gas irradiated by X-rays.  We note here that this mosaic corresponds to the integrated fluorescent emission over a time period of $\sim$20 years, with observations in the GC and CMZ being deeper and spanning a longer time period than those of the inner Galactic plane where the observations are generally shallower ($\sim$20\,ks) and span a time period of four years. Moreover, different parts of the map were observed at different times. We observe that the reflection emission is concentrated in the central degrees of the GC along the CMZ. We show a zoomed-in view of the central $\sim$2$\times$1 degrees in the bottom panel of Fig.\,\ref{fig.cleanmap}. In this panel, bright sources masked out during the fitting process and the profile extraction are indicated by solid white circles. Known molecular cloud positions are outlined with dashed white regions, while the location of Sgr\,A$^{*}$ is marked by an X. All indicated regions show 6.4\,keV emission with particular intensity at the Sgr A molecular complex.

\begin{figure*}[!htbp]
\centering
\includegraphics[angle=0,trim=0cm 0cm 0cm 0cm,width=1.8\columnwidth]{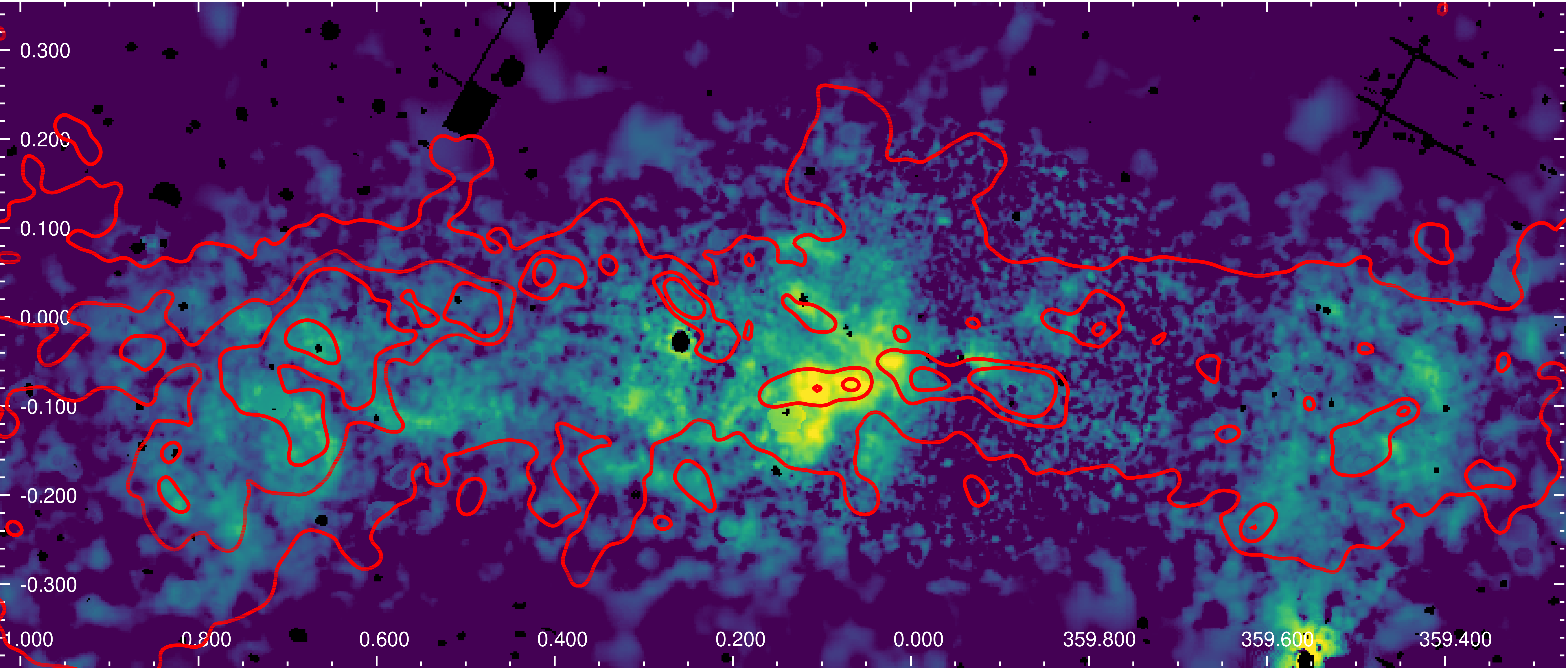}
\includegraphics[angle=0,trim=0cm 0cm 0cm 0cm,width=1.8\columnwidth]{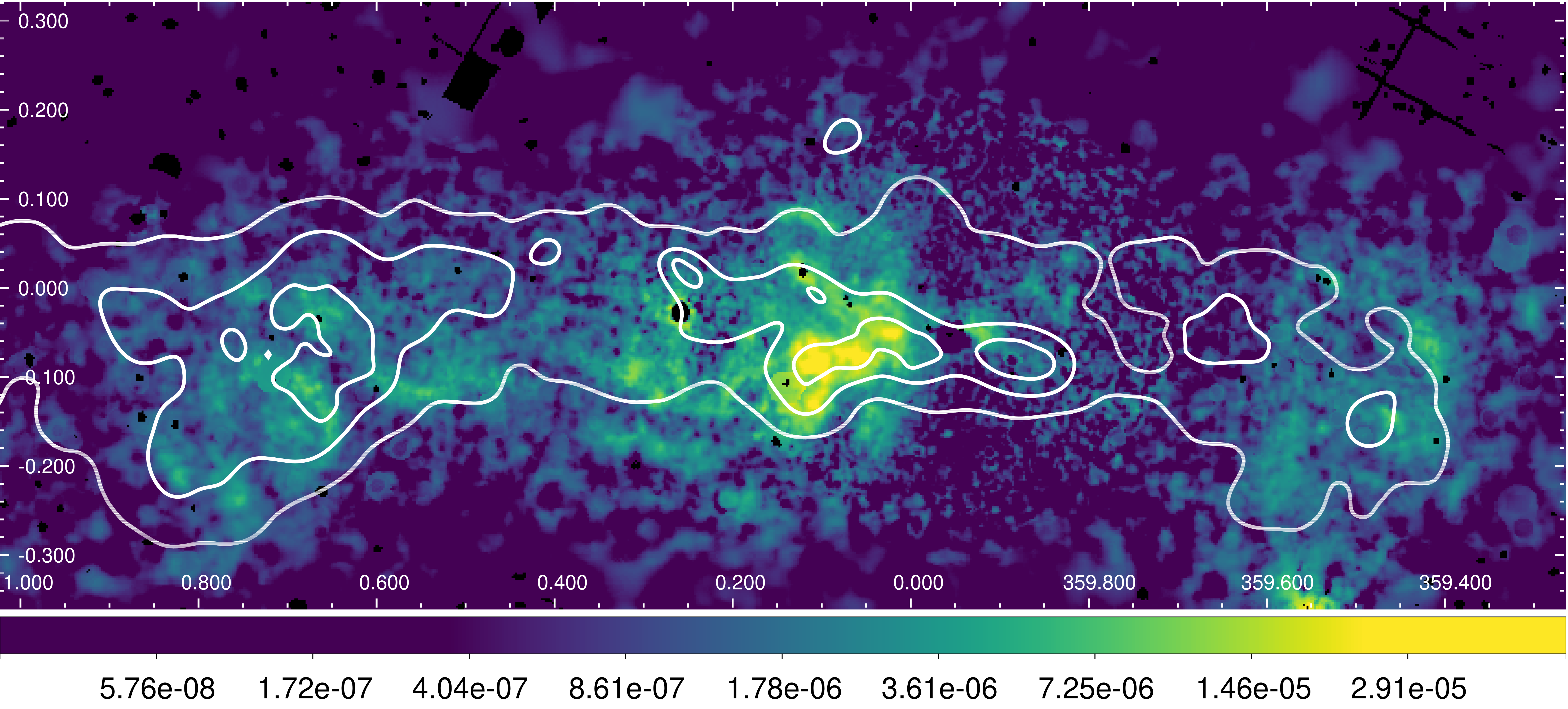}
\caption{X64 cleaned map with overlaid contours of the cold and dense molecular gas in the CMZ. Top: The \herschel{} SPIRE 250$\mu$m emission contours \citep{molinari11}. Bottom: The $N_{2}H^{+}$ molecular emission contours \citep{jones12}.}
\label{fig.x64coldgas}
\end{figure*}

A valuable test to investigate how the cleaned X64 map correlates with reflected emission from molecular clouds is to compare its distribution along the CMZ with tracers of molecular gas. For this purpose, we use contours from two surveys that outline molecular gas emission in the Galactic centre. Specifically, we use the \herschel{} SPIRE 250$\mu$m map from the Hi-GAL survey Key Project \citep{molinari10,molinari11}, which traces a continuous chain of cold and dense clumps, and the N2H$^{+}$ molecular gas emission map from the Mopra survey of the CMZ \citep{jones12}.
In Fig.\,\ref{fig.x64coldgas}, we overlay contours from \citet{molinari11} and \citet{jones12} on the cleaned X64 map, shown in the top and bottom panels, respectively. This comparison reveals a strong correlation with the distribution of cold and dense clumps as well as  the molecular gas, suggesting that this emission indeed arises from X-rays reflected by cold gas in molecular clouds illuminated by past X-ray radiation. Perfect agreement between the distribution of cold gas and the reflected X-ray emission is not expected due to the uneven illumination of the clouds and the different observation times by \xmm{}.

\begin{figure*}[!htbp]
\centering
\includegraphics[angle=0,trim=0cm 0cm 0cm 0cm,width=1.8\columnwidth]{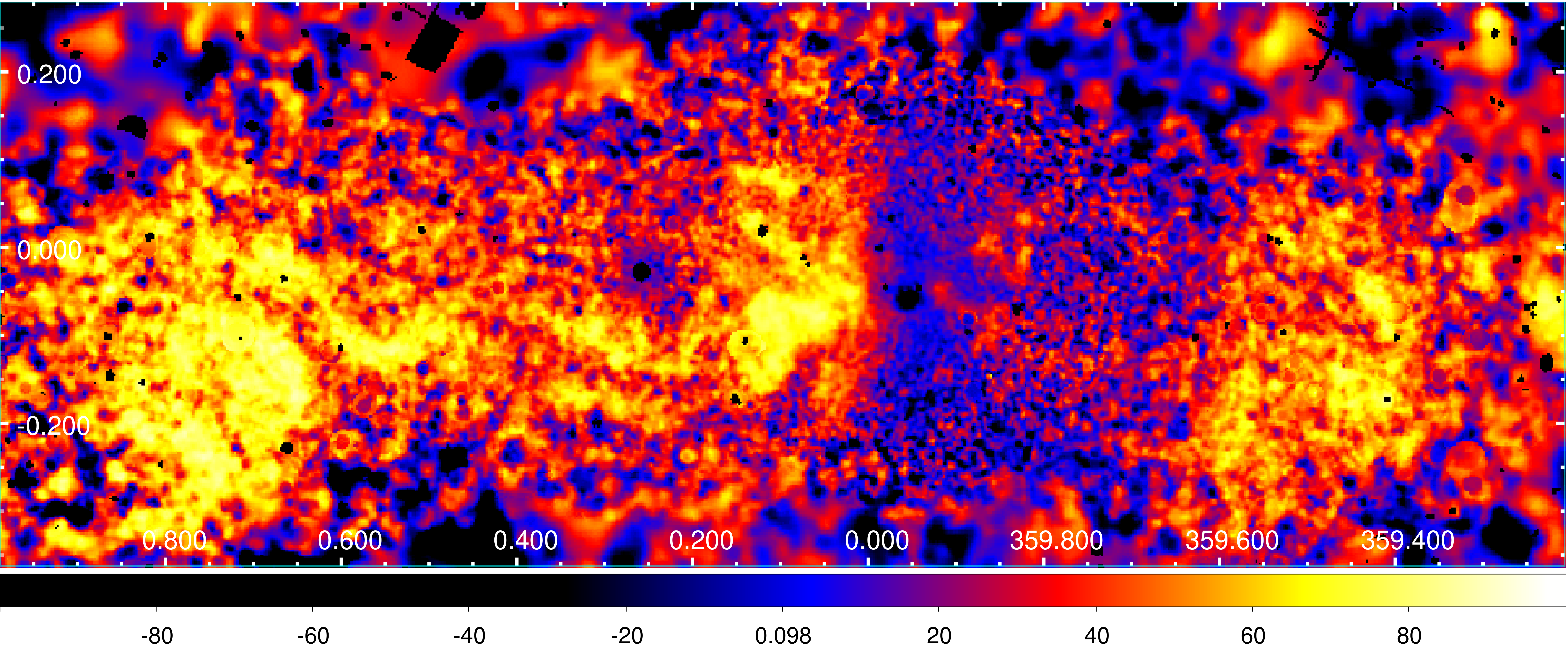}
\caption{
Fraction of the flux in the X64 band that remains after
the best-fitting contributions to this band from stars and continuum
have been removed. The remaining flux can plausibly be attributed to the
fluorescent emission from molecular gas.
} 
\label{fig.percentage}
\end{figure*}

In Fig.\,\ref{fig.percentage} we show the fraction of the  original (raw) 6.4\,keV map that remains after the removal of stars and continuum, plausibly attributed to clean fluorescent emission.
We notice that at the position of the most prominent fluorescent features, more than 80\% of the raw emission (yellow features) could be attributed to  fluorescent emission from molecular gas. Negative regions (shown in black) are present, but of significantly smaller amplitude
comparable to the noise level, while areas of the map that do not show reflection remain close to zero (shown in blue). Using the fiducial model, we estimate the "non-reflection" contribution to the 6.4\,keV band, within a central elliptical region consistent to the CMZ (Fig.\,\ref{fig.examples}),  to a median value of 65\%.
For regions outside the CMZ the "clean reflection" component approaches zero as is also evident by the latitudinal and longitudinal profiles (Fig.\,\ref{fig.profiles}). The contribution of the non-reflection component to the CMZ region is significant, and should be considered in future analyses that take into account the pure reflection emission.
In addition, the cleaned map can also be used to set upper limits on the 6.4\,keV flux from unilluminated molecular complexes, which is crucial for understanding the low-energy cosmic ray ionization rate and other steady sources of reflection-like emission. 

In this work, we filter out the contribution of the diffuse component that affects the 6.4\,keV band while also correlating with the projected mass distribution. As a key point for future studies, we emphasize that the cleaned X64 map serves as a proxy for the time-averaged reflection emission, which is closely linked to the constant illumination of molecular gas in the CMZ. This illumination could be driven by Sgr\,A$^{*}$, cosmic rays, or a multitude of bright X-ray sources in the Galactic Center region, each predicting different correlations with molecular gas density. For example, single sources (e.g., Sgr A* or XRBs) would show a distance-dependent trend, cosmic ray models might correlate with gamma-ray emission morphologies, and in the case of multiple X-ray sources, we expect a mix of uniform illumination and a centrally peaked contribution from the NSC.

\section{Conclusions}\label{conclusions}

In this work, we utilized mosaics of all available \xmm{} observations for the Galactic Center and inner disc, along with stellar mass distribution 
 models of our Galaxy, to decompose the 6.4\,keV line emission into truly reflected emission and contributions from other physical components. Using the spatial correlations of the continuum (X50), 6.7\,keV (X67), and 6.4\,keV (X64) bands, together with stellar mass distribution maps, we derived a clean fluorescent X64 map. Key findings include:

\begin{itemize}

\item For the 6.30-6.50 keV band (X64), a linear combination of the NSC+NSD and bar+Disc maps effectively accounts for much of the diffuse emission that is traced by the distribution of stars.
This reveals that outside the CMZ region, most of the 6.4 keV emission can be attributed to unresolved point sources.

\item A linear combination of the X67 and X50 maps produces even cleaner maps (in terms of the residual noise in the X64 map). This, on the one hand, confirms that the diffuse continuum emission correlates strongly with the distribution of stars \citep[i.e.][]{revnivtsev07}, for instance similar to the NSD and bar components. In addition, using these maps reduces the residuals associated with the brightest compact sources.

\item Masking only the bright sources before fitting the maps is insufficient to avoid negative residuals in the final cleaned maps as the high density of sources in the GC region likely breaks the assumption that physically unrelated signals are spatially uncorrelated, which essential for using our method. Applying an entire elliptical mask to exclude bright emission in the Galactic Center yields optimal results with minimal or no negative residuals, as demonstrated in the Galactic longitudinal and latitudinal profiles.

\item Our fiducial clean X64 map reveals that the truly diffuse observed emission is confined to the CMZ region. 

\item The cleaned map shows a strong correlation with the distribution of molecular gas as well as dense and cold gas in the Galactic Center. This component likely represents X-rays reflected by cold gas in molecular clouds illuminated by X-rays.

\end{itemize}

Given that the  "residual" 6.4\,keV emission map we derive, closely aligns with the distribution of molecular and cold and dense gas along the CMZ, we argue that this map offers the best estimate for the X-ray reflection signal (observed with \xmm{} over last two decades).
Furthermore, we estimate that approximately 65\% of the ridge emission contributes to the observed 6.4\,keV emission in the CMZ. This contribution should be taken into account in future studies, including polarization analyses of the reflected X-ray continuum from molecular clouds and statistical examinations of reflection surface brightness fluctuations.

\section*{Data availability}

The cleaned map presented in Fig.\,\ref{fig.cleanmap} is available in electronic form at the CDS via anonymous ftp to cdsarc.cds.unistra.fr (130.79.128.5) or via \href{ https://cdsarc.cds.unistra.fr/cgi-bin/qcat?J/A+A/}{ https://cdsarc.cds.unistra.fr/cgi-bin/qcat?J/A+A/}.
  
\begin{acknowledgements}
KA acknowledges support from \chandra{} grants GO3-24033B, GO0-21010X, and TM9-20001X, and \textit{JWST} grant JWST-GO-01905.002-A. 
IK acknowledges support by the COMPLEX project from the European Research Council (ERC) under the European Union’s Horizon 2020 research and innovation program grant agreement ERC-2019-AdG 882679.
MCS acknowledges financial support from the European Research Council under the ERC Starting Grant ``GalFlow'' (grant 101116226) and from Fondazione Cariplo under the grant ERC attrattivit\`{a} n. 2023-3014.
GP acknowledges support from the European Research Council (ERC) under the European Union’s Horizon 2020 research and innovation program HotMilk (grant agreement No. 865637), support from Bando per il Finanziamento della Ricerca Fondamentale 2022 dell’Istituto Nazionale di Astrofisica (INAF): GO Large program and from the Framework per l’Attrazione e il Rafforzamento delle Eccellenze (FARE) per la ricerca in Italia (R20L5S39T9). 
We also made use of NASA’s Astrophysics Data System Bibliographic Services.
\end{acknowledgements}




\bibliographystyle{aa} 
\bibliography{aa54389-25} 

\begin{thebibliography}{78}
\expandafter\ifx\csname natexlab\endcsname\relax\def\natexlab#1{#1}\fi

\bibitem[{{Anastasopoulou} {et~al.}(2023){Anastasopoulou}, {Ponti}, {Sormani},
  {Locatelli}, {Haberl}, {Morris}, {Churazov}, {Sch{\"o}del}, {Maitra},
  {Campana}, {Di Teodoro}, {Jin}, {Khabibullin}, {Mondal}, {Sasaki}, {Zhang},
  \& {Zheng}}]{anastasopoulou23}
{Anastasopoulou}, K., {Ponti}, G., {Sormani}, M.~C., {et~al.} 2023, \aap, 671,
  A55

\bibitem[{{Bally} {et~al.}(1988){Bally}, {Stark}, {Wilson}, \&
  {Henkel}}]{bally88}
{Bally}, J., {Stark}, A.~A., {Wilson}, R.~W., \& {Henkel}, C. 1988, \apj, 324,
  223

\bibitem[{{Barret} {et~al.}(2000){Barret}, {Olive}, {Boirin}, {Done},
  {Skinner}, \& {Grindlay}}]{barret2000}
{Barret}, D., {Olive}, J.~F., {Boirin}, L., {et~al.} 2000, \apj, 533, 329

\bibitem[{{Bland-Hawthorn} \& {Gerhard}(2016)}]{Bland-Hawthorn2016}
{Bland-Hawthorn}, J. \& {Gerhard}, O. 2016, \araa, 54, 529

\bibitem[{{Bykov}(2002)}]{bykov02}
{Bykov}, A.~M. 2002, \aap, 390, 327

\bibitem[{{Capelli} {et~al.}(2011){Capelli}, {Warwick}, {Porquet}, {Gillessen},
  \& {Predehl}}]{capelli11}
{Capelli}, R., {Warwick}, R.~S., {Porquet}, D., {Gillessen}, S., \& {Predehl},
  P. 2011, \aap, 530, A38

\bibitem[{{Capelli} {et~al.}(2012){Capelli}, {Warwick}, {Porquet}, {Gillessen},
  \& {Predehl}}]{capelli12}
{Capelli}, R., {Warwick}, R.~S., {Porquet}, D., {Gillessen}, S., \& {Predehl},
  P. 2012, \aap, 545, A35

\bibitem[{{Chatzopoulos} {et~al.}(2015){Chatzopoulos}, {Fritz}, {Gerhard},
  {Gillessen}, {Wegg}, {Genzel}, \& {Pfuhl}}]{chatzopoulos15}
{Chatzopoulos}, S., {Fritz}, T.~K., {Gerhard}, O., {et~al.} 2015, \mnras, 447,
  948

\bibitem[{{Chernyshov} {et~al.}(2018){Chernyshov}, {Ko}, {Krivonos}, {Dogiel},
  \& {Cheng}}]{chernyshov18}
{Chernyshov}, D.~O., {Ko}, C.~M., {Krivonos}, R.~A., {Dogiel}, V.~A., \&
  {Cheng}, K.~S. 2018, \apj, 863, 85

\bibitem[{{Chuard} {et~al.}(2018){Chuard}, {Terrier}, {Goldwurm}, {Clavel},
  {Soldi}, {Morris}, {Ponti}, {Walls}, \& {Chernyakova}}]{chuard18}
{Chuard}, D., {Terrier}, R., {Goldwurm}, A., {et~al.} 2018, \aap, 610, A34

\bibitem[{{Churazov} {et~al.}(2017{\natexlab{a}}){Churazov}, {Khabibullin},
  {Ponti}, \& {Sunyaev}}]{churazov17}
{Churazov}, E., {Khabibullin}, I., {Ponti}, G., \& {Sunyaev}, R.
  2017{\natexlab{a}}, \mnras, 468, 165

\bibitem[{{Churazov} {et~al.}(2017{\natexlab{b}}){Churazov}, {Khabibullin},
  {Sunyaev}, \& {Ponti}}]{churazov17a}
{Churazov}, E., {Khabibullin}, I., {Sunyaev}, R., \& {Ponti}, G.
  2017{\natexlab{b}}, \mnras, 465, 45

\bibitem[{{Churazov} {et~al.}(2017{\natexlab{c}}){Churazov}, {Khabibullin},
  {Sunyaev}, \& {Ponti}}]{2017MNRAS.465...45C}
{Churazov}, E., {Khabibullin}, I., {Sunyaev}, R., \& {Ponti}, G.
  2017{\natexlab{c}}, \mnras, 465, 45

\bibitem[{{Clavel} {et~al.}(2014){Clavel}, {Soldi}, {Terrier}, {Tatischeff},
  {Maurin}, {Ponti}, {Goldwurm}, \& {Decourchelle}}]{clavel14}
{Clavel}, M., {Soldi}, S., {Terrier}, R., {et~al.} 2014, \mnras, 443, L129

\bibitem[{{Clavel} {et~al.}(2013){Clavel}, {Terrier}, {Goldwurm}, {Morris},
  {Ponti}, {Soldi}, \& {Trap}}]{clavel13}
{Clavel}, M., {Terrier}, R., {Goldwurm}, A., {et~al.} 2013, \aap, 558, A32

\bibitem[{{Cooke} {et~al.}(1969){Cooke}, {Griffiths}, \& {Pounds}}]{cooke69}
{Cooke}, B.~A., {Griffiths}, R.~E., \& {Pounds}, K.~A. 1969, \nat, 224, 134

\bibitem[{{Do} {et~al.}(2018){Do}, {Kerzendorf}, {Konopacky}, {Marcinik},
  {Ghez}, {Lu}, \& {Morris}}]{do18}
{Do}, T., {Kerzendorf}, W., {Konopacky}, Q., {et~al.} 2018, \apjl, 855, L5

\bibitem[{{Dogiel} {et~al.}(2009){Dogiel}, {Cheng}, {Chernyshov}, {Bamba},
  {Ichimura}, {Inoue}, {Ko}, {Kokubun}, {Maeda}, {Mitsuda}, \&
  {Yamasaki}}]{dogiel09}
{Dogiel}, V., {Cheng}, K.-S., {Chernyshov}, D., {et~al.} 2009, \pasj, 61, 901

\bibitem[{{Feldmeier-Krause} {et~al.}(2017){Feldmeier-Krause}, {Kerzendorf},
  {Neumayer}, {Sch{\"o}del}, {Nogueras-Lara}, {Do}, {de Zeeuw}, \&
  {Kuntschner}}]{feldmeierkrause17}
{Feldmeier-Krause}, A., {Kerzendorf}, W., {Neumayer}, N., {et~al.} 2017,
  \mnras, 464, 194

\bibitem[{{Fritz} {et~al.}(2021){Fritz}, {Patrick}, {Feldmeier-Krause},
  {Sch{\"o}del}, {Schultheis}, {Gerhard}, {Nandakumar}, {Neumayer},
  {Nogueras-Lara}, \& {Prieto}}]{fritz21}
{Fritz}, T.~K., {Patrick}, L.~R., {Feldmeier-Krause}, A., {et~al.} 2021, \aap,
  649, A83

\bibitem[{{Gravity Collaboration} {et~al.}(2019){Gravity Collaboration},
  {Abuter}, {Amorim}, {Baub{\"o}ck}, {Berger}, {Bonnet}, {Brandner},
  {Cl{\'e}net}, {Coud{\'e} Du Foresto}, {de Zeeuw}, {Dexter}, {Duvert},
  {Eckart}, {Eisenhauer}, {F{\"o}rster Schreiber}, {Garcia}, {Gao}, {Gendron},
  {Genzel}, {Gerhard}, {Gillessen}, {Habibi}, {Haubois}, {Henning}, {Hippler},
  {Horrobin}, {Jim{\'e}nez-Rosales}, {Jocou}, {Kervella}, {Lacour},
  {Lapeyr{\`e}re}, {Le Bouquin}, {L{\'e}na}, {Ott}, {Paumard}, {Perraut},
  {Perrin}, {Pfuhl}, {Rabien}, {Rodriguez Coira}, {Rousset}, {Scheithauer},
  {Sternberg}, {Straub}, {Straubmeier}, {Sturm}, {Tacconi}, {Vincent}, {von
  Fellenberg}, {Waisberg}, {Widmann}, {Wieprecht}, {Wiezorrek}, {Woillez}, \&
  {Yazici}}]{gravity19}
{Gravity Collaboration}, {Abuter}, R., {Amorim}, A., {et~al.} 2019, \aap, 625,
  L10

\bibitem[{{Heard} \& {Warwick}(2013)}]{heard13sources}
{Heard}, V. \& {Warwick}, R.~S. 2013, \mnras, 428, 3462

\bibitem[{{Inui} {et~al.}(2009){Inui}, {Koyama}, {Matsumoto}, \&
  {Tsuru}}]{inui09}
{Inui}, T., {Koyama}, K., {Matsumoto}, H., \& {Tsuru}, T.~G. 2009, \pasj, 61,
  S241

\bibitem[{{Jin} {et~al.}(2017){Jin}, {Ponti}, {Haberl}, \& {Smith}}]{jin2017}
{Jin}, C., {Ponti}, G., {Haberl}, F., \& {Smith}, R. 2017, \mnras, 468, 2532

\bibitem[{{Jin} {et~al.}(2018){Jin}, {Ponti}, {Haberl}, {Smith}, \&
  {Valencic}}]{jin18}
{Jin}, C., {Ponti}, G., {Haberl}, F., {Smith}, R., \& {Valencic}, L. 2018,
  \mnras, 477, 3480

\bibitem[{{Jones} {et~al.}(2012){Jones}, {Burton}, {Cunningham},
  {Requena-Torres}, {Menten}, {Schilke}, {Belloche}, {Leurini},
  {Mart{\'\i}n-Pintado}, {Ott}, \& {Walsh}}]{jones12}
{Jones}, P.~A., {Burton}, M.~G., {Cunningham}, M.~R., {et~al.} 2012, \mnras,
  419, 2961

\bibitem[{{Kallman} \& {White}(1989)}]{kallman89}
{Kallman}, T. \& {White}, N.~E. 1989, \apj, 341, 955

\bibitem[{{Khabibullin} {et~al.}(2022){Khabibullin}, {Churazov}, \&
  {Sunyaev}}]{khabibullin22}
{Khabibullin}, I., {Churazov}, E., \& {Sunyaev}, R. 2022, \mnras, 509, 6068

\bibitem[{{Koyama}(2018)}]{koyama18}
{Koyama}, K. 2018, \pasj, 70, R1

\bibitem[{{Koyama} {et~al.}(1989){Koyama}, {Awaki}, {Kunieda}, {Takano}, \&
  {Tawara}}]{koyama89}
{Koyama}, K., {Awaki}, H., {Kunieda}, H., {Takano}, S., \& {Tawara}, Y. 1989,
  \nat, 339, 603

\bibitem[{{Koyama} {et~al.}(2007){Koyama}, {Inui}, {Hyodo}, {Matsumoto},
  {Tsuru}, {Maeda}, {Murakami}, {Yamauchi}, {Kissel}, {Chan}, \&
  {Soong}}]{koyama07b}
{Koyama}, K., {Inui}, T., {Hyodo}, Y., {et~al.} 2007, \pasj, 59, 221

\bibitem[{{Koyama} {et~al.}(1996){Koyama}, {Maeda}, {Sonobe}, {Takeshima},
  {Tanaka}, \& {Yamauchi}}]{koyama96}
{Koyama}, K., {Maeda}, Y., {Sonobe}, T., {et~al.} 1996, \pasj, 48, 249

\bibitem[{{Koyama} {et~al.}(1986){Koyama}, {Makishima}, {Tanaka}, \&
  {Tsunemi}}]{koyama86}
{Koyama}, K., {Makishima}, K., {Tanaka}, Y., \& {Tsunemi}, H. 1986, \pasj, 38,
  121

\bibitem[{{Krivonos} {et~al.}(2017){Krivonos}, {Clavel}, {Hong}, {Mori},
  {Ponti}, {Poutanen}, {Rahoui}, {Tomsick}, \& {Tsygankov}}]{krivonos17}
{Krivonos}, R., {Clavel}, M., {Hong}, J., {et~al.} 2017, \mnras, 468, 2822

\bibitem[{{Krivonos} {et~al.}(2007){Krivonos}, {Revnivtsev}, {Churazov},
  {Sazonov}, {Grebenev}, \& {Sunyaev}}]{krivonos07}
{Krivonos}, R., {Revnivtsev}, M., {Churazov}, E., {et~al.} 2007, \aap, 463, 957

\bibitem[{{Krivonos} {et~al.}(2025){Krivonos}, {Shtykovskaya}, \&
  {Sazonov}}]{krivonos25}
{Krivonos}, R., {Shtykovskaya}, E., \& {Sazonov}, S. 2025, Journal of High
  Energy Astrophysics, 45, 96

\bibitem[{{Kuznetsova} {et~al.}(2019){Kuznetsova}, {Krivonos}, {Clavel},
  {Lutovinov}, {Chernyshov}, {Hong}, {Mori}, {Ponti}, {Tomsick}, \&
  {Zhang}}]{kuznetsova19}
{Kuznetsova}, E., {Krivonos}, R., {Clavel}, M., {et~al.} 2019, \mnras, 484,
  1627

\bibitem[{{Kuznetsova} {et~al.}(2022){Kuznetsova}, {Krivonos}, {Lutovinov}, \&
  {Clavel}}]{kuznetsova22}
{Kuznetsova}, E., {Krivonos}, R., {Lutovinov}, A., \& {Clavel}, M. 2022,
  \mnras, 509, 1605

\bibitem[{{Launhardt} {et~al.}(2002){Launhardt}, {Zylka}, \&
  {Mezger}}]{launhardt02}
{Launhardt}, R., {Zylka}, R., \& {Mezger}, P.~G. 2002, \aap, 384, 112

\bibitem[{{Marin} {et~al.}(2023){Marin}, {Churazov}, {Khabibullin},
  {Ferrazzoli}, {Di Gesu}, {Barnouin}, {Di Marco}, {Middei}, {Vikhlinin},
  {Costa}, {Soffitta}, {Muleri}, {Sunyaev}, {Forman}, {Kraft}, {Bianchi},
  {Donnarumma}, {Petrucci}, {Enoto}, {Agudo}, {Antonelli}, {Bachetti},
  {Baldini}, {Baumgartner}, {Bellazzini}, {Bongiorno}, {Bonino}, {Brez},
  {Bucciantini}, {Capitanio}, {Castellano}, {Cavazzuti}, {Chen}, {Ciprini}, {De
  Rosa}, {Del Monte}, {Di Lalla}, {Doroshenko}, {Dovciak}, {Ehlert},
  {Evangelista}, {Fabiani}, {Garcia}, {Gunji}, {Hayashida}, {Heyl}, {Ingram},
  {Iwakiri}, {Jorstad}, {Kaaret}, {Karas}, {Kitaguchi}, {Kolodziejczak},
  {Krawczynski}, {La Monaca}, {Latronico}, {Liodakis}, {Maldera}, {Manfreda},
  {Marinucci}, {Marscher}, {Marshall}, {Massaro}, {Matt}, {Mitsuishi},
  {Mizuno}, {Negro}, {Ng}, {O'Dell}, {Omodei}, {Oppedisano}, {Papitto},
  {Pavlov}, {Peirson}, {Perri}, {Pesce-Rollins}, {Pilia}, {Possenti},
  {Poutanen}, {Puccetti}, {Ramsey}, {Rankin}, {Ratheesh}, {Roberts}, {Romani},
  {Sgr{\`o}}, {Slane}, {Spandre}, {Swartz}, {Tamagawa}, {Tavecchio}, {Taverna},
  {Tawara}, {Tennant}, {Thomas}, {Tombesi}, {Trois}, {Tsygankov}, {Turolla},
  {Vink}, {Weisskopf}, {Wu}, {Xie}, \& {Zane}}]{marin23}
{Marin}, F., {Churazov}, E., {Khabibullin}, I., {et~al.} 2023, arXiv e-prints,
  arXiv:2304.06967

\bibitem[{{Molinari} {et~al.}(2011){Molinari}, {Bally}, {Noriega-Crespo},
  {Compi{\`e}gne}, {Bernard}, {Paradis}, {Martin}, {Testi}, {Barlow}, {Moore},
  {Plume}, {Swinyard}, {Zavagno}, {Calzoletti}, {Di Giorgio}, {Elia},
  {Faustini}, {Natoli}, {Pestalozzi}, {Pezzuto}, {Piacentini}, {Polenta},
  {Polychroni}, {Schisano}, {Traficante}, {Veneziani}, {Battersby}, {Burton},
  {Carey}, {Fukui}, {Li}, {Lord}, {Morgan}, {Motte}, {Schuller},
  {Stringfellow}, {Tan}, {Thompson}, {Ward-Thompson}, {White}, \&
  {Umana}}]{molinari11}
{Molinari}, S., {Bally}, J., {Noriega-Crespo}, A., {et~al.} 2011, \apjl, 735,
  L33

\bibitem[{{Molinari} {et~al.}(2010){Molinari}, {Swinyard}, {Bally}, {Barlow},
  {Bernard}, {Martin}, {Moore}, {Noriega-Crespo}, {Plume}, {Testi}, {Zavagno},
  {Abergel}, {Ali}, {Anderson}, {Andr{\'e}}, {Baluteau}, {Battersby},
  {Beltr{\'a}n}, {Benedettini}, {Billot}, {Blommaert}, {Bontemps}, {Boulanger},
  {Brand}, {Brunt}, {Burton}, {Calzoletti}, {Carey}, {Caselli}, {Cesaroni},
  {Cernicharo}, {Chakrabarti}, {Chrysostomou}, {Cohen}, {Compiegne}, {de
  Bernardis}, {de Gasperis}, {di Giorgio}, {Elia}, {Faustini}, {Flagey},
  {Fukui}, {Fuller}, {Ganga}, {Garcia-Lario}, {Glenn}, {Goldsmith}, {Griffin},
  {Hoare}, {Huang}, {Ikhenaode}, {Joblin}, {Joncas}, {Juvela}, {Kirk},
  {Lagache}, {Li}, {Lim}, {Lord}, {Marengo}, {Marshall}, {Masi}, {Massi},
  {Matsuura}, {Minier}, {Miville-Desch{\^e}nes}, {Montier}, {Morgan}, {Motte},
  {Mottram}, {M{\"u}ller}, {Natoli}, {Neves}, {Olmi}, {Paladini}, {Paradis},
  {Parsons}, {Peretto}, {Pestalozzi}, {Pezzuto}, {Piacentini}, {Piazzo},
  {Polychroni}, {Pomar{\`e}s}, {Popescu}, {Reach}, {Ristorcelli}, {Robitaille},
  {Robitaille}, {Rod{\'o}n}, {Roy}, {Royer}, {Russeil}, {Saraceno}, {Sauvage},
  {Schilke}, {Schisano}, {Schneider}, {Schuller}, {Schulz}, {Sibthorpe},
  {Smith}, {Smith}, {Spinoglio}, {Stamatellos}, {Strafella}, {Stringfellow},
  {Sturm}, {Taylor}, {Thompson}, {Traficante}, {Tuffs}, {Umana}, {Valenziano},
  {Vavrek}, {Veneziani}, {Viti}, {Waelkens}, {Ward-Thompson}, {White},
  {Wilcock}, {Wyrowski}, {Yorke}, \& {Zhang}}]{molinari10}
{Molinari}, S., {Swinyard}, B., {Bally}, J., {et~al.} 2010, \aap, 518, L100

\bibitem[{{Muno} {et~al.}(2004){Muno}, {Baganoff}, {Bautz}, {Feigelson},
  {Garmire}, {Morris}, {Park}, {Ricker}, \& {Townsley}}]{muno04}
{Muno}, M.~P., {Baganoff}, F.~K., {Bautz}, M.~W., {et~al.} 2004, \apj, 613, 326

\bibitem[{{Muno} {et~al.}(2007){Muno}, {Baganoff}, {Brandt}, {Park}, \&
  {Morris}}]{muno07}
{Muno}, M.~P., {Baganoff}, F.~K., {Brandt}, W.~N., {Park}, S., \& {Morris},
  M.~R. 2007, \apjl, 656, L69

\bibitem[{{Murakami} {et~al.}(2001){Murakami}, {Koyama}, \&
  {Maeda}}]{murakami01}
{Murakami}, H., {Koyama}, K., \& {Maeda}, Y. 2001, \apj, 558, 687

\bibitem[{{Neumayer} {et~al.}(2020){Neumayer}, {Seth}, \&
  {B{\"o}ker}}]{neumayer20}
{Neumayer}, N., {Seth}, A., \& {B{\"o}ker}, T. 2020, \aapr, 28, 4

\bibitem[{{Nishiyama} {et~al.}(2013){Nishiyama}, {Yasui}, {Nagata},
  {Yoshikawa}, {Uchiyama}, {Sch{\"o}del}, {Hatano}, {Sato}, {Sugitani},
  {Suenaga}, {Kwon}, \& {Tamura}}]{nishiyama13}
{Nishiyama}, S., {Yasui}, K., {Nagata}, T., {et~al.} 2013, \apjl, 769, L28

\bibitem[{{Park} {et~al.}(2004){Park}, {Muno}, {Baganoff}, {Maeda}, {Morris},
  {Howard}, {Bautz}, \& {Garmire}}]{park04}
{Park}, S., {Muno}, M.~P., {Baganoff}, F.~K., {et~al.} 2004, \apj, 603, 548

\bibitem[{{Planck Collaboration} {et~al.}(2016){Planck Collaboration},
  {Aghanim}, {Arnaud}, {Ashdown}, {Aumont}, {Baccigalupi}, {Banday},
  {Barreiro}, {Bartlett}, {Bartolo}, {Battaner}, {Battye}, {Benabed},
  {Beno{\^\i}t}, {Benoit-L{\'e}vy}, {Bernard}, {Bersanelli}, {Bielewicz},
  {Bock}, {Bonaldi}, {Bonavera}, {Bond}, {Borrill}, {Bouchet}, {Burigana},
  {Butler}, {Calabrese}, {Cardoso}, {Catalano}, {Challinor}, {Chiang},
  {Christensen}, {Churazov}, {Clements}, {Colombo}, {Combet}, {Comis},
  {Coulais}, {Crill}, {Curto}, {Cuttaia}, {Danese}, {Davies}, {Davis}, {de
  Bernardis}, {de Rosa}, {de Zotti}, {Delabrouille}, {D{\'e}sert}, {Dickinson},
  {Diego}, {Dolag}, {Dole}, {Donzelli}, {Dor{\'e}}, {Douspis}, {Ducout},
  {Dupac}, {Efstathiou}, {Elsner}, {En{\ss}lin}, {Eriksen}, {Fergusson},
  {Finelli}, {Forni}, {Frailis}, {Fraisse}, {Franceschi}, {Frejsel},
  {Galeotta}, {Galli}, {Ganga}, {G{\'e}nova-Santos}, {Giard},
  {Gonz{\'a}lez-Nuevo}, {G{\'o}rski}, {Gregorio}, {Gruppuso}, {Gudmundsson},
  {Hansen}, {Harrison}, {Henrot-Versill{\'e}}, {Hern{\'a}ndez-Monteagudo},
  {Herranz}, {Hildebrandt}, {Hivon}, {Holmes}, {Hornstrup}, {Huffenberger},
  {Hurier}, {Jaffe}, {Jones}, {Juvela}, {Keih{\"a}nen}, {Keskitalo}, {Kneissl},
  {Knoche}, {Kunz}, {Kurki-Suonio}, {Lacasa}, {Lagache}, {L{\"a}hteenm{\"a}ki},
  {Lamarre}, {Lasenby}, {Lattanzi}, {Leonardi}, {Lesgourgues}, {Levrier},
  {Liguori}, {Lilje}, {Linden-V{\o}rnle}, {L{\'o}pez-Caniego},
  {Mac{\'\i}as-P{\'e}rez}, {Maffei}, {Maggio}, {Maino}, {Mandolesi},
  {Mangilli}, {Maris}, {Martin}, {Mart{\'\i}nez-Gonz{\'a}lez}, {Masi},
  {Matarrese}, {Melchiorri}, {Melin}, {Migliaccio}, {Miville-Desch{\^e}nes},
  {Moneti}, {Montier}, {Morgante}, {Mortlock}, {Munshi}, {Murphy}, {Naselsky},
  {Nati}, {Natoli}, {Noviello}, {Novikov}, {Novikov}, {Paci}, {Pagano},
  {Pajot}, {Paoletti}, {Pasian}, {Patanchon}, {Perdereau}, {Perotto},
  {Pettorino}, {Piacentini}, {Piat}, {Pierpaoli}, {Pietrobon}, {Plaszczynski},
  {Pointecouteau}, {Polenta}, {Ponthieu}, {Pratt}, {Prunet}, {Puget}, {Rachen},
  {Reinecke}, {Remazeilles}, {Renault}, {Renzi}, {Ristorcelli}, {Rocha},
  {Rossetti}, {Roudier}, {Rubi{\~n}o-Mart{\'\i}n}, {Rusholme}, {Sandri},
  {Santos}, {Sauv{\'e}}, {Savelainen}, {Savini}, {Scott}, {Spencer},
  {Stolyarov}, {Stompor}, {Sunyaev}, {Sutton}, {Suur-Uski}, {Sygnet}, {Tauber},
  {Terenzi}, {Toffolatti}, {Tomasi}, {Tramonte}, {Tristram}, {Tucci},
  {Tuovinen}, {Valenziano}, {Valiviita}, {Van Tent}, {Vielva}, {Villa}, {Wade},
  {Wandelt}, {Wehus}, {Yvon}, {Zacchei}, \& {Zonca}}]{2016A&A...594A..22P}
{Planck Collaboration}, {Aghanim}, N., {Arnaud}, M., {et~al.} 2016, \aap, 594,
  A22

\bibitem[{{Ponti} {et~al.}(2019){Ponti}, {Hofmann}, {Churazov}, {Morris},
  {Haberl}, {Nandra}, {Terrier}, {Clavel}, \& {Goldwurm}}]{ponti19}
{Ponti}, G., {Hofmann}, F., {Churazov}, E., {et~al.} 2019, \nat, 567, 347

\bibitem[{{Ponti} {et~al.}(2013){Ponti}, {Morris}, {Terrier}, \&
  {Goldwurm}}]{ponti13}
{Ponti}, G., {Morris}, M.~R., {Terrier}, R., \& {Goldwurm}, A. 2013, in
  Astrophysics and Space Science Proceedings, Vol.~34, Cosmic Rays in
  Star-Forming Environments, ed. D.~F. {Torres} \& O.~{Reimer}, 331

\bibitem[{{Ponti} {et~al.}(2015){Ponti}, {Morris}, {Terrier}, {Haberl},
  {Sturm}, {Clavel}, {Soldi}, {Goldwurm}, {Predehl}, {Nandra}, {B{\'e}langer},
  {Warwick}, \& {Tatischeff}}]{ponti15}
{Ponti}, G., {Morris}, M.~R., {Terrier}, R., {et~al.} 2015, \mnras, 453, 172

\bibitem[{{Ponti} {et~al.}(2010){Ponti}, {Terrier}, {Goldwurm}, {Belanger}, \&
  {Trap}}]{ponti10}
{Ponti}, G., {Terrier}, R., {Goldwurm}, A., {Belanger}, G., \& {Trap}, G. 2010,
  \apj, 714, 732

\bibitem[{{Portail} {et~al.}(2017){Portail}, {Gerhard}, {Wegg}, \&
  {Ness}}]{portail17}
{Portail}, M., {Gerhard}, O., {Wegg}, C., \& {Ness}, M. 2017, \mnras, 465, 1621

\bibitem[{{Revnivtsev} {et~al.}(2006{\natexlab{a}}){Revnivtsev}, {Molkov}, \&
  {Sazonov}}]{revnivtsev06b}
{Revnivtsev}, M., {Molkov}, S., \& {Sazonov}, S. 2006{\natexlab{a}}, \mnras,
  373, L11

\bibitem[{{Revnivtsev} {et~al.}(2009){Revnivtsev}, {Sazonov}, {Churazov},
  {Forman}, {Vikhlinin}, \& {Sunyaev}}]{revnivtsev09}
{Revnivtsev}, M., {Sazonov}, S., {Churazov}, E., {et~al.} 2009, \nat, 458, 1142

\bibitem[{{Revnivtsev} {et~al.}(2006{\natexlab{b}}){Revnivtsev}, {Sazonov},
  {Gilfanov}, {Churazov}, \& {Sunyaev}}]{revnivtsev06a}
{Revnivtsev}, M., {Sazonov}, S., {Gilfanov}, M., {Churazov}, E., \& {Sunyaev},
  R. 2006{\natexlab{b}}, \aap, 452, 169

\bibitem[{{Revnivtsev} {et~al.}(2007){Revnivtsev}, {Vikhlinin}, \&
  {Sazonov}}]{revnivtsev07}
{Revnivtsev}, M., {Vikhlinin}, A., \& {Sazonov}, S. 2007, \aap, 473, 857

\bibitem[{{Ryu} {et~al.}(2013){Ryu}, {Nobukawa}, {Nakashima}, {Tsuru},
  {Koyama}, \& {Uchiyama}}]{ryu13}
{Ryu}, S.~G., {Nobukawa}, M., {Nakashima}, S., {et~al.} 2013, \pasj, 65, 33

\bibitem[{{Sch{\"o}del} {et~al.}(2014){Sch{\"o}del}, {Feldmeier}, {Kunneriath},
  {Stolovy}, {Neumayer}, {Amaro-Seoane}, \& {Nishiyama}}]{schoedel14}
{Sch{\"o}del}, R., {Feldmeier}, A., {Kunneriath}, D., {et~al.} 2014, \aap, 566,
  A47

\bibitem[{{Schultheis} {et~al.}(2021){Schultheis}, {Fritz}, {Nandakumar},
  {Rojas-Arriagada}, {Nogueras-Lara}, {Feldmeier-Krause}, {Gerhard},
  {Neumayer}, {Patrick}, {Prieto}, {Sch{\"o}del}, {Mastrobuono-Battisti}, \&
  {Sormani}}]{schulteis21}
{Schultheis}, M., {Fritz}, T.~K., {Nandakumar}, G., {et~al.} 2021, \aap, 650,
  A191

\bibitem[{{Sormani} {et~al.}(2022{\natexlab{a}}){Sormani}, {Gerhard},
  {Portail}, {Vasiliev}, \& {Clarke}}]{sormani22bar}
{Sormani}, M.~C., {Gerhard}, O., {Portail}, M., {Vasiliev}, E., \& {Clarke}, J.
  2022{\natexlab{a}}, \mnras, 514, L1

\bibitem[{{Sormani} {et~al.}(2022{\natexlab{b}}){Sormani}, {Sanders}, {Fritz},
  {Smith}, {Gerhard}, {Sch{\"o}del}, {Magorrian}, {Neumayer}, {Nogueras-Lara},
  {Feldmeier-Krause}, {Mastrobuono-Battisti}, {Schultheis}, {Shahzamanian},
  {Vasiliev}, {Klessen}, {Lucas}, \& {Minniti}}]{sormani22nsd}
{Sormani}, M.~C., {Sanders}, J.~L., {Fritz}, T.~K., {et~al.}
  2022{\natexlab{b}}, \mnras, 512, 1857

\bibitem[{{Stel} {et~al.}(2025){Stel}, {Ponti}, {Haardt}, \&
  {Sormani}}]{stel25}
{Stel}, G., {Ponti}, G., {Haardt}, F., \& {Sormani}, M. 2025, \aap, 695, A52

\bibitem[{{Sunyaev} {et~al.}(1993){Sunyaev}, {Markevitch}, \&
  {Pavlinsky}}]{sunyaev93}
{Sunyaev}, R.~A., {Markevitch}, M., \& {Pavlinsky}, M. 1993, \apj, 407, 606

\bibitem[{{Terrier} {et~al.}(2010){Terrier}, {Ponti}, {B{\'e}langer},
  {Decourchelle}, {Tatischeff}, {Goldwurm}, {Trap}, {Morris}, \&
  {Warwick}}]{terrier10}
{Terrier}, R., {Ponti}, G., {B{\'e}langer}, G., {et~al.} 2010, \apj, 719, 143

\bibitem[{{Uchiyama} {et~al.}(2011){Uchiyama}, {Nobukawa}, {Tsuru}, {Koyama},
  \& {Matsumoto}}]{uchiyama11}
{Uchiyama}, H., {Nobukawa}, M., {Tsuru}, T., {Koyama}, K., \& {Matsumoto}, H.
  2011, \pasj, 63, S903

\bibitem[{{Valinia} {et~al.}(2000){Valinia}, {Tatischeff}, {Arnaud}, {Ebisawa},
  \& {Ramaty}}]{valinia00}
{Valinia}, A., {Tatischeff}, V., {Arnaud}, K., {Ebisawa}, K., \& {Ramaty}, R.
  2000, \apj, 543, 733

\bibitem[{{Wang} {et~al.}(2006){Wang}, {Dong}, \& {Lang}}]{wang06}
{Wang}, Q.~D., {Dong}, H., \& {Lang}, C. 2006, \mnras, 371, 38

\bibitem[{{Wang} {et~al.}(2002){Wang}, {Gotthelf}, \& {Lang}}]{wang02}
{Wang}, Q.~D., {Gotthelf}, E.~V., \& {Lang}, C.~C. 2002, \nat, 415, 148

\bibitem[{{Worrall} {et~al.}(1982){Worrall}, {Marshall}, {Boldt}, \&
  {Swank}}]{worrall82}
{Worrall}, D.~M., {Marshall}, F.~E., {Boldt}, E.~A., \& {Swank}, J.~H. 1982,
  \apj, 255, 111

\bibitem[{{Xu} {et~al.}(2016){Xu}, {Wang}, \& {Li}}]{xu16}
{Xu}, X.-j., {Wang}, Q.~D., \& {Li}, X.-D. 2016, \apj, 818, 136

\bibitem[{{Yamauchi} \& {Koyama}(1993)}]{yamauchi93}
{Yamauchi}, S. \& {Koyama}, K. 1993, \apj, 404, 620

\bibitem[{{Yamauchi} {et~al.}(2016){Yamauchi}, {Nobukawa}, {Nobukawa},
  {Uchiyama}, \& {Koyama}}]{yamauchi16}
{Yamauchi}, S., {Nobukawa}, K.~K., {Nobukawa}, M., {Uchiyama}, H., \& {Koyama},
  K. 2016, \pasj, 68, 59

\bibitem[{{Yuasa} {et~al.}(2012){Yuasa}, {Makishima}, \& {Nakazawa}}]{yuasa12}
{Yuasa}, T., {Makishima}, K., \& {Nakazawa}, K. 2012, \apj, 753, 129

\bibitem[{{Yusef-Zadeh} {et~al.}(2002){Yusef-Zadeh}, {Law}, \&
  {Wardle}}]{yused-zadeh02}
{Yusef-Zadeh}, F., {Law}, C., \& {Wardle}, M. 2002, \apjl, 568, L121

\bibitem[{{Yusef-Zadeh} {et~al.}(2007){Yusef-Zadeh}, {Muno}, {Wardle}, \&
  {Lis}}]{yusef-zadeh07}
{Yusef-Zadeh}, F., {Muno}, M., {Wardle}, M., \& {Lis}, D.~C. 2007, \apj, 656,
  847

\bibitem[{{Zhu} {et~al.}(2018){Zhu}, {Li}, \& {Morris}}]{zhu18}
{Zhu}, Z., {Li}, Z., \& {Morris}, M.~R. 2018, \apjs, 235, 26

\end{thebibliography}



\begin{appendix}

\onecolumn

\section{\xmm{} recent observations}
\begin{longtable}{lrrrrrr}
\caption{\label{tab.xmm}\xmm{} observations}\\
\toprule
OBSID& Exp pn  &  Exp MOS1  &  Exp MOS2 & clean Exp pn  &  clean Exp MOS1  &  clean Exp MOS2    \\
&sec&sec & sec& sec &sec &sec\\
\endfirsthead
\caption{continued.}\\
\bottomrule
OBSID& Exp pn  &  Exp MOS1  &  Exp MOS2 & clean Exp pn  & clean Exp MOS1  & clean Exp MOS2  \\
&sec&sec & sec& sec &sec &sec \\
\midrule
\endhead
\hline
\endfoot
\hline
0886021101 & 26137 &  29746 &  29688 &  11100 &  18086 &  17766   \\    
0886021701 & 26054 &  28161 &  28143 &  13400 &  23709 &  24832   \\    
0886031101 & 28771 &  31658 &  31988 &  7800  &  22929 &  26340   \\    
0886031601 & 22746 &  24592 &  24586 &  18155 &  24614 &  24586   \\    
0886040101 & 20738 &  22591 &  22529 &  16388 &  19433 &  20624   \\    
0886040201 & 20853 &  22705 &  22685 &  16157 &  20378 &  21346   \\    
0886040301 & 22032 &  24608 &  24599 &  21534 &  24556 &  24599   \\    
0886040401 & 19747 &  21619 &  21580 &  15750 &  21619 &  21599   \\    
0886040501 & 24530 &  29962 &  29755 &  12800 &  17469 &  18540   \\    
0886040601 & 27769 &  29933 &  25319 &  12641 &  20017 &  17255   \\    
0886040701 & 25462 &  27319 &  27300 &  12000 &  22877 &  25248   \\    
0886061201 & 19725 &  21573 &  21554 &  3262  &  9971  &  12740   \\    
0886070301 & 18520 &  23575 &  23505 &  13626 &  15182 &  15847   \\    
0886070401 & 24757 &  26620 &  26601 &  19159 &  23970 &  25273   \\    
0886070501 & 19758 &  20191 &  19259 &  19759 &  20191 &  19259   \\    
0886070601 & 23480 &  26620 &  26583 &  23480 &  26464 &  26497   \\    
0886070701 & 19747 &  21584 &  21554 &  18755 &  21619 &  21599   \\    
0886070801 & 28459 &  30321 &  30301 &  24060 &  28452 &  28920   \\    
0886070901 & 27684 &  30006 &  29999 &  19457 &  25651 &  26827   \\    
0886071001 & 19756 &  21619 &  21599 &  19759 &  21619 &  21599   \\    
0886071101 & 24160 &  26020 &  26000 &  24160 &  26020 &  26000   \\    
0886071201 & 20550 &  22408 &  22386 &  20553 &  22418 &  22399   \\    
0886071301 & 22961 &  25985 &  25956 &  22962 &  26020 &  26000   \\    
0886080101 & 24389 &  29555 &  29333 &  16309 &  22357 &  23480   \\    
0886080201 & 19759 &  21619 &  21599 &  19762 &  21619 &  21599   \\    
0886080301 & 24781 &  25996 &  25390 &  15163 &  19293 &  19357   \\    
0886080401 & 18758 &  20607 &  20600 &  18759 &  20620 &  20600   \\    
0886080501 & 29530 &  31612 &  31601 &  20060 &  26007 &  26373   \\    
0886080601 & 19756 &  21606 &  21599 &  19756 &  21619 &  21599   \\    
0886080701 & 29762 &  31620 &  31601 &  29762 &  31620 &  31601   \\    
0886080801 & 19761 &  21619 &  21599 &  19762 &  21619 &  21599   \\    
0886080901 & 22653 &  29225 &  29328 &  14312 &  19585 &  19840   \\    
0886081001 & 24737 &  26595 &  26575 &  9737  &  17079 &  19451   \\    
0886081101 & 19762 &  21619 &  21599 &  19762 &  21619 &  21599   \\    
0886081201 & 19762 &  21619 &  21599 &  19762 &  21619 &  21599   \\    
0886081301 & 20758 &  22621 &  22597 &  20762 &  22621 &  22597   \\    
0886090201 & 14295 &  16522 &  16500 &  10200 &  14141 &  14276   \\    
0886090301 & 19691 &  21514 &  21476 &  17595 &  21341 &  21523   \\    
0886090401 & 19754 &  21606 &  21586 &  16862 &  21567 &  21599   \\    
0886090501 & 11260 &  13109 &  13099 &  8760  &  12797 &  13099   \\    
0886090601 & 19761 &  21608 &  21575 &  15861 &  21151 &  21443   \\    
0886090701 & 8252  &  10098 &  10070 &  6156  &  10120 &  10099   \\    
0886090801 & 19755 &  21592 &  21583 &  19655 &  21608 &  21599   \\    
0886090901 & 18254 &  20110 &  20085 &  18263 &  20121 &  20098   \\    
0886091001 & 8233  &  10077 &  10070 &  8236  &  10087 &  10078   \\    
0886091101 & 9271  &  14813 &  14721 &  0     &  1456  &  1612    \\    
0886091201 & 8262  &  10120 &  10099 &  8262  &  10120 &  10099   \\    
0886100101 & 11762 &  13608 &  13598 &  11462 &  13619 &  13598   \\    
0886100201 & 10762 &  12620 &  12600 &  10562 &  12620 &  12600   \\    
0886100301 & 8262  &  10120 &  10099 &  8262  &  10120 &  10099   \\    
0886100401 & 11947 &  13792 &  13798 &  10656 &  13485 &  13642   \\    
0886100501 & 19759 &  21619 &  21599 &  4861  &  11271 &  11615   \\    
0886100601 & 16017 &  19373 &  19357 &  0     &  12685 &  16812   \\    
0886100701 & 3261  &  1404  &  359   &  3261  &  1404  &  359     \\    
0886100801 & 8255  &  10109 &  10086 &  8255  &  10120 &  10099   \\    
0886100901 & 8262  &  10120 &  10099 &  8262  &  10120 &  10099   \\    
0886101001 & 19754 &  21619 &  21583 &  12059 &  16211 &  15827   \\    
0886101201 & 21241 &  28559 &  28551 &  19760 &  9669  &  16097   \\    
0886110101 & 19710 &  21565 &  21528 &  19714 &  21579 &  21565   \\    
0886110301 & 19760 &  21608 &  21583 &  15661 &  21619 &  21599   \\    
0886110401 & 8262  &  10120 &  10099 &  5962  &  9513  &  9683    \\    
0886110501 & 24756 &  26609 &  26601 &  19363 &  25366 &  25457   \\    
0886110601 & 19756 &  20461 &  19443 &  14762 &  20472 &  19443   \\    
0886120901 & 18850 &  21598 &  21599 &  14052 &  21502 &  21495   \\    
0886121001 & 19755 &  21619 &  21586 &  18259 &  21411 &  21599   \\    
0886121101 & 8262  &  10120 &  10099 &  7862  &  10120 &  10099   \\    
0886121201 & 24647 &  26620 &  26598 &  12900 &  17936 &  18694   \\    
0886121301 & 24702 &  26606 &  26601 &  22559 &  24802 &  24805   \\    
0916790101 & 13250 &  15099 &  15101 &  2960  &  8987  &  10444   \\    
0916790201 & 8262  &  10106 &  10088 &  8262  &  10120 &  10099   \\    
0916790301 & 8262  &  10120 &  10099 &  8262  &  10120 &  10099   \\    
0916790401 & 8259  &  10120 &  10083 &  8259  &  10120 &  10099   \\    
0916790501 & 8262  &  10120 &  10099 &  8262  &  10120 &  10099   \\    
0916790601 & 8258  &  10120 &  10099 &  8260  &  10120 &  10099   \\    
0916790701 & 8254  &  10109 &  10099 &  8254  &  10109 &  10099   \\    
0916790801 & 8256  &  10120 &  10099 &  8262  &  10120 &  10099   \\    
0916790901 & 18259 &  20121 &  20088 &  18159 &  20121 &  20101   \\    
0916791001 & 8262  &  10120 &  10099 &  8262  &  9912  &  9527    \\    
0916791101 & 9161  &  11019 &  11001 &  9061  &  7069  &  5957    \\    
091679120  & 18262 &  10120 &  10099 &    0   &   5762 &     0 	  \\     
091679130  & 16407 &  10120 &  10099 &    0   &  1683  &     0 	  \\     
0916800201 & 15059 &  16900 &  16900 &  15060 &  16908 &  16900   \\    
0916800301 & 8254  &  10109 &  10099 &  8254  &  10120 &  10099   \\    
0916800401 & 8253  &  10106 &  10099 &  8258  &  10120 &  10099   \\    
0916800501 & 8252  &  10112 &  10073 &  8255  &  10120 &  10099   \\    
0916800601 & 13207 &  18895 &  18874 &  7912  &  12997 &  13549   \\    
0916800701 & 9862  &  11721 &  11700 &  9862  &  11721 &  11700   \\    
0916800801 & 8262  &  10120 &  10099 &  8262  &  10120 &  10099   \\    
0916800901 & 8261  &  10120 &  10099 &  8261  &  10120 &  10099   \\    
0916801001 & 8261  &  10120 &  10099 &  8261  &  10120 &  10099   \\    
0916801101 & 8262  &  10120 &  10099 &  6262  &  9758  &  9915    \\    
0916801201 & 7056  &  10101 &  10083 &  5097  &  9860  &  10047   \\    
0916801301 & 7924  &  18922 &  18782 &  2942  &  5665  &  5803    \\    
0916801401 & 18255 &  20110 &  20101 &  13859 &  19705 &  19944   \\    
0916801501 & 8261  &  10120 &  10099 &  8261  &  10120 &  10099   \\    
0916810101 & 8262  &  10120 &  10099 &  8262  &  10120 &  10099   \\    
0916810201 & 8262  &  10120 &  10099 &  8262  &  10120 &  10099   \\    
0916810301 & 8262  &  10120 &  10099 &  8262  &  10120 &  10099   \\    
0916810401 & 8251  &  10109 &  10083 &  8254  &  10120 &  10091   \\    
0916810501 & 16741 &  18587 &  18585 &  16746 &  18601 &  18585   \\    
0916810601 & 10812 &  18263 &  18249 &  0     &  7485  &  8036    \\    
0916810701 & 8262  &  10120 &  10099 &  3862  &  10120 &  10099   \\    
0916810801 & 3570  &  1596  &  348   &  3570  &  1596  &  348     \\    
0916810901 & 8261  &  10120 &  10099 &  4162  &  10120 &  10099   \\    
0916811001 & 2240  &  2187  &  2915  &  2240  &  2187  &  2915    \\    
0916811201 & 15880 &  18871 &  18850 &  9600  &  15493 &  16304   \\    
0916811401 & 7682  &  10425 &  10408 &  7282  &  10444 &  10421   \\    
0916811501 & 13270 &  18177 &  18156 &  7756  &  10776 &  10509   \\    
0916811701 & 8251  &  10109 &  10086 &  7953  &  10120 &  10099   \\    
0916811801 & 8261  &  10120 &  10099 &  8262  &  10120 &  10099   \\    
0916811901 & 9883  &  13308 &  13239 &  4187  &  8586  &  8598    \\    
0932190101 & 7659  &  11011 &  11001 &  3258  &  4002  &  4159    \\    
0932190201 & 8262  &  10120 &  10099 &  2800  &  6792  &  8227    \\    
0932190301 & 8262  &  10120 &  10099 &  3962  &  8768  &  9163    \\    
0932190401 & 8261  &  10120 &  10099 &  3961  &  9860  &  10099   \\    
0932190501 & 8759  &  10619 &  10593 &  5659  &  10515 &  10600   \\    
0932190601 & 9762  &  11621 &  11599 &  9062  &  11621 &  11599   \\    
0932190701 & 9761  &  11621 &  11599 &  9661  &  11621 &  11599   \\    
0932190801 & 9752  &  11599 &  11588 &  8954  &  11621 &  11599   \\    
0932190901 & 9740  &  11591 &  11583 &  8282  &  10430 &  10532   \\    
0932191001 & 7292  &  10120 &  10099 &  7292  &  10120 &  10099   \\    
0932191101 & 8262  &  10120 &  10099 &  8262  &  10120 &  10099   \\    
0932200101 & 8262  &  10120 &  10099 &  8262  &  10120 &  10099   \\    
0932200201 & 8253  &  10112 &  10073 &  8255  &  10120 &  10099   \\    
0932200301 & 8259  &  10120 &  10099 &  7359  &  9964  &  9891    \\    
0932200401 & 8262  &  10120 &  10099 &  8262  &  10120 &  10099   \\    
0932200501 & 8254  &  10104 &  10083 &  8255  &  10120 &  10099   \\    
0932200601 & 23063 &  28850 &  28923 &  8198  &  14505 &  17604   \\    
0932200701 & 21470 &  23609 &  23585 &  18462 &  21696 &  21726   \\    
0932200801 & 19762 &  21619 &  21599 &  19762 &  21619 &  21599   \\    
0932200901 & 8262  &  10120 &  10099 &  362   &  7936  &  9943    \\    
0932201001 & 7893  &  10120 &  10099 &  1900  &  9028  &  9787   \\    
0934200101 & 21300&  21650 & 21646&  15711 &  17567 &  17090  \\ 
\end{longtable}

\section{Limitations of the bright source masking strategy}\label{brightmask}

In this Section, we present the results obtained when masking only the bright sources. We also explain why this strategy does not perform as well as masking, in addition, the entire central region affected by reflection using an elliptical mask. The latter is the approach adopted in our main analysis (Section\,\ref{contaminatigcomponents}; Fig.\,\ref{fig.examples}).

\begin{figure*}[!htbp]
\centering
\includegraphics[angle=0,trim=0cm 0cm 0cm 0cm,width=0.99\columnwidth]{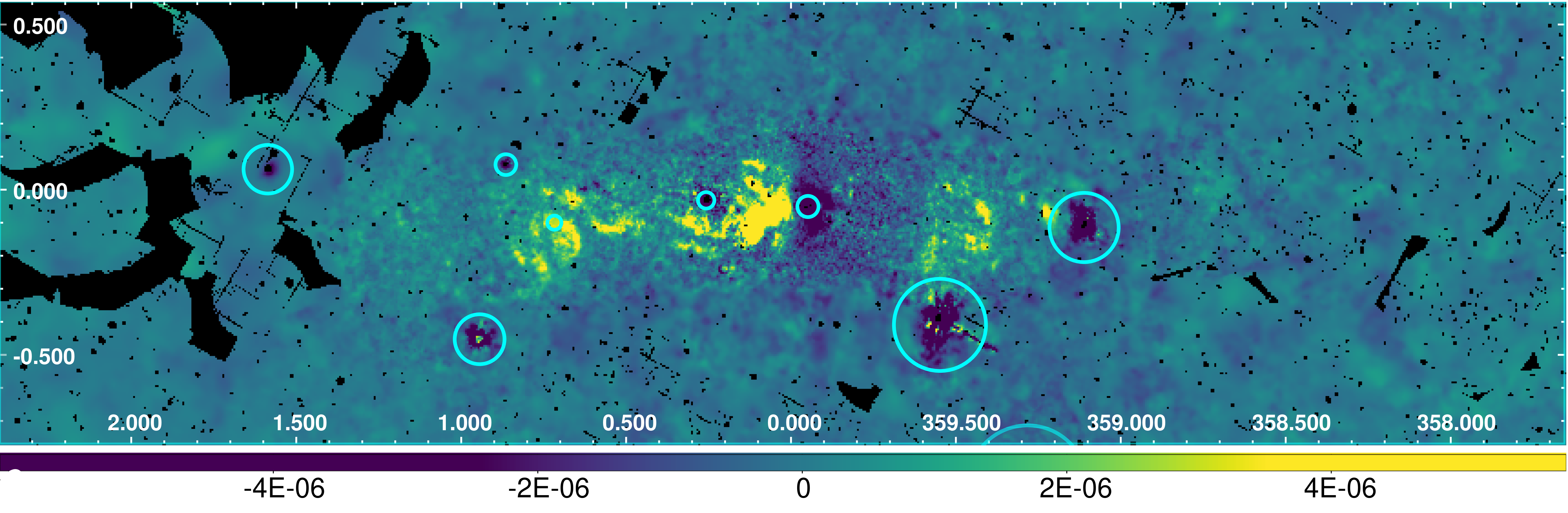}
\caption{Example of the cleaned X64 band maps (X64 minus C-X67-X50) when masking only for bright sources. The mask is shown with the cyan circles. Significant negative residuals remain at the Galactic centre near Sgr\,A$^{*}$ and near the bight sources.} 
\label{fig.examplebrightmask}
\end{figure*}

\begin{figure*}[!htbp]
 	\centering	   \includegraphics[width=0.7\columnwidth]{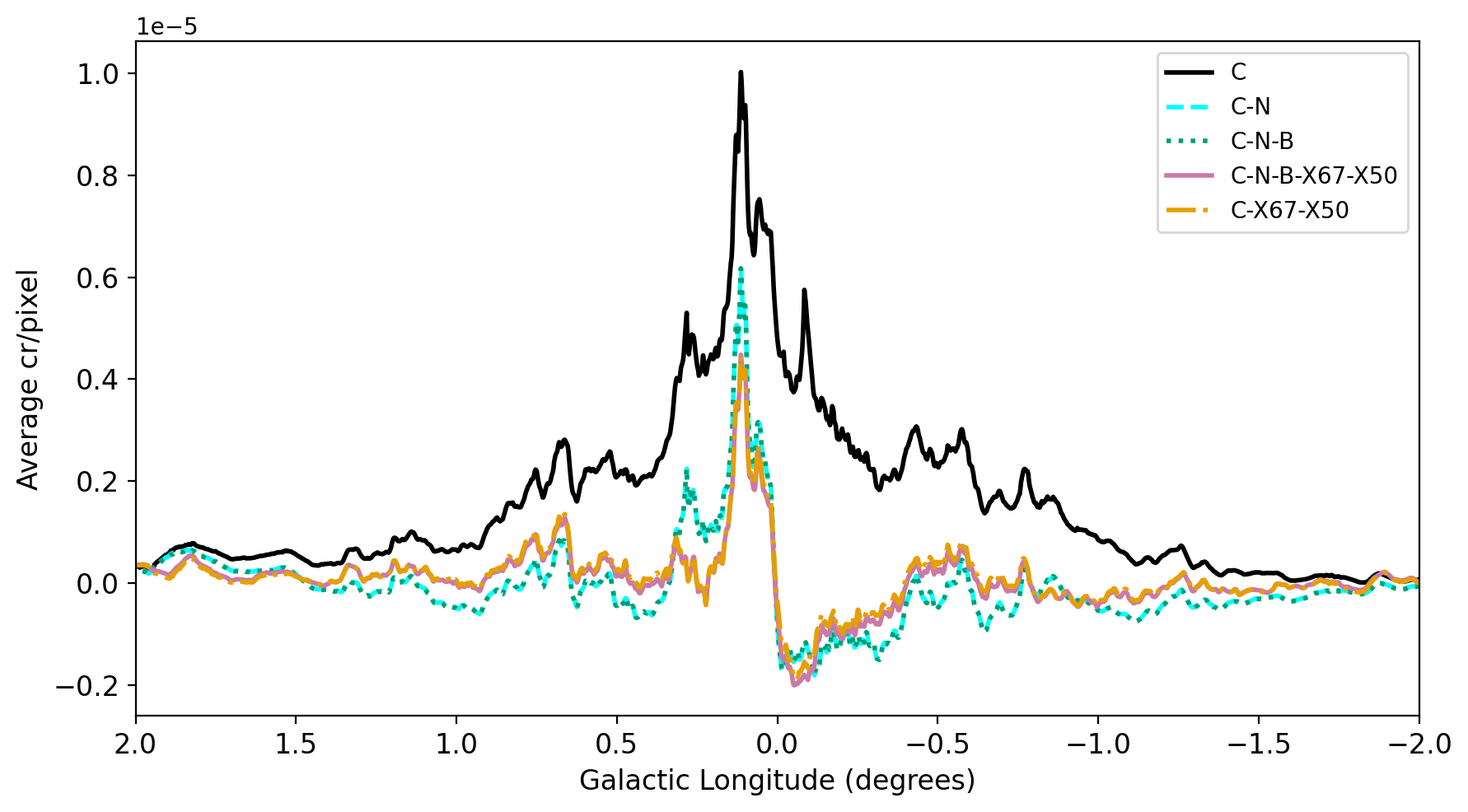}
  \caption{Longitudinal profiles of the X64 cleaned map when masking only for bright sources. Significant negative residuals remain particularly at the Galactic centre near Sgr\,A$^{*}$.}
  \label{fig.longprofilesbright}
  \end{figure*}

The cleaned map (Fig.\,\ref{fig.examplebrightmask}) using only the bright source mask shows significant negative residuals when minimising the variance across the entire map—particularly in the Galactic Centre near Sgr\,A$^{*}$ and around bright sources.
This likely happens as both the peak of the reflected emission and the thermal emission are concentrated in the Galactic Centre. Hence, some of the reflected emission is removed. This is further reflected in the Galactic longitudinal profiles. When using the bright source mask (Fig.\,\ref{fig.longprofilesbright}), all models show significant negative residuals particularly close to the GC. When comparing these results to those obtained with elliptical masking (Figs.\,\ref{fig.examples} and \ref{fig.profiles}), the improved performance of the adopted method, with almost no negative residuals, becomes evident.

\end{appendix}
\end{document}